\documentclass[aps,prd,preprint,12pt,superscriptaddress,nofootinbib,a4paper]{revtex4-1}

\usepackage[utf8]{inputenc}
\usepackage{amsmath,amssymb}
\usepackage[separate-uncertainty=true]{siunitx}
\usepackage{graphicx}
\usepackage[usenames,dvipsnames]{xcolor}
\usepackage[caption=false]{subfig}
\usepackage{hyperref}
\usepackage[capitalize]{cleveref}
\usepackage{booktabs}
\usepackage{tabularx}
\usepackage{xspace}
\usepackage[normalem]{ulem}

\usepackage[compat=1.1.0]{tikz-feynman}

\allowdisplaybreaks

\AtBeginDocument{
\heavyrulewidth=.08em
\lightrulewidth=.05em
\cmidrulewidth=.03em
\belowrulesep=.65ex
\belowbottomsep=0pt
\aboverulesep=.4ex
\abovetopsep=0pt
\cmidrulesep=\doublerulesep
\cmidrulekern=.5em
\defaultaddspace=.5em

\newcolumntype{C}{>{\centering\arraybackslash}X}
\newcolumntype{b}{C}
\newcolumntype{s}{>{\hsize=.6\hsize}C}
\newcolumntype{R}{>{\raggedleft\arraybackslash}X}
}

\graphicspath{{graphics/}}

\newcommand{\eqdot}{\,.}
\newcommand{\eqcomma}{\,,}

\newcommand{\ie}{i.e.~}
\newcommand{\eg}{e.g.~}

\newcommand{\Ztwo}{\ensuremath{{\mathbb{Z}_2}}\xspace}
\newcommand{\HiggsBounds}{\texttt{HiggsBounds}\xspace}
\newcommand{\HiggsSignals}{\texttt{HiggsSignals}\xspace}
\newcommand{\BP}[1]{\textbf{BP#1}}
\newcommand{\BR}[1]{\ensuremath{\text{BR}(#1)}}

\DeclareSIUnit{\pb}{pb}
\DeclareSIUnit{\fb}{fb}

\begin{document}

\date{\today}
\title{{\Large Two-real-scalar-singlet extension of the SM:\texorpdfstring{\\[0.2cm]}{} LHC phenomenology and benchmark scenarios}}
\author{Tania Robens}
\email{trobens@irb.hr}
\affiliation{Ruder Boskovic Institute, Bijenicka cesta 54, 10000 Zagreb, Croatia}
\author{Tim Stefaniak}
\email{tim.stefaniak@desy.de}
\affiliation{DESY, Notkestraße 85, 22607 Hamburg, Germany}
\author{Jonas Wittbrodt}
\email{jonas.wittbrodt@desy.de}
\affiliation{DESY, Notkestraße 85, 22607 Hamburg, Germany}

\preprint{DESY 19-142}

\renewcommand{\abstractname}{\texorpdfstring{\vspace{0.5cm}}{} Abstract}

\begin{abstract}
    \vspace{0.5cm}
    We investigate the LHC phenomenology of a model where the Standard Model
    (SM) scalar sector is extended by two real scalar singlets. A
    $\Ztwo\otimes\Ztwo'$ discrete symmetry is imposed to reduce the number of
    scalar potential parameters, which is spontaneously broken by the vacuum
    expectation values of the singlet fields. As a result, all three neutral
    scalar fields mix, leading to three neutral CP-even scalar bosons, out of
    which one is identified with the observed Higgs boson at $\SI{125}{\GeV}$.
    We explore all relevant collider signatures of the three scalars in this
    model. Besides the single production of a  scalar boson decaying directly to
    SM particle final states, we extensively discuss the possibility of resonant
    multi-scalar production. The latter includes decays of the produced scalar
    boson to two identical scalars (``symmetric decays''), as well as to two
    different scalars (``asymmetric decays''). Furthermore, we discuss the
    possibility of successive decays to the lightest scalar states (``cascade
    decays''), which lead to experimentally spectacular three- and four-Higgs
    final states. We provide six benchmark scenarios for detailed experimental
    studies of these Higgs-to-Higgs decay signatures.
\end{abstract}

\maketitle

\section{Introduction}

The Large Hadron Collider (LHC) at CERN is the first experimental facility that
directly probes the mechanism of electroweak symmetry breaking (EWSB),
described in the Standard Model of particle physics (SM) by the
Brout-Englert-Higgs
mechanism~\cite{Englert:1964et,Higgs:1964ia,Higgs:1964pj,Guralnik:1964eu,Higgs:1966ev,Kibble:1967sv}.
The milestone discovery of a Higgs boson with a mass of $\sim \SI{125}{\GeV}$
in 2012~\cite{Aad:2012tfa,Chatrchyan:2012xdj} and the ongoing measurements of
its properties at the LHC~(see \eg Ref.~\cite{Khachatryan:2016vau}) open the
door to a deeper understanding of the structure of EWSB\@. Indeed, this
experimental endeavor may reveal first signs of new physics beyond the SM
(BSM), as many well-motivated BSM extensions affect the phenomenology of the
observed scalar particle. However, by the end of Run-II of the LHC, with the
full collected data of $\sim \SI{150}{\fb^{-1}}$ per experiment (ATLAS and CMS)
still being analyzed, Higgs signal rate measurements in various production and
decay channels~\cite{ Sirunyan:2017elk,
    Sirunyan:2017dgc,  CMS:1900lgv, CMS-PAS-HIG-18-030,Sirunyan:2018hbu, Sirunyan:2018shy,Sirunyan:2018egh, CMS:2018dmv, Aad:2019mbh, CMS:2019chr,CMS:2019pyn} are
so far in very good agreement with the SM predictions.

Extensions of the SM by scalar singlets are
among the simplest possible model beyond the SM (BSM). The most general
extension of the SM by $n$ real scalar singlet fields $\phi_i$
($i\in[1,\ldots,n]$) has a scalar potential of the form

\begin{equation}
    V(\phi_i, \Phi) =V_{\text{singlets}} (\phi_i,\Phi) + V_\text{SM}(\Phi)\eqcomma
    \label{eq:generalpotential}
\end{equation}
where
\begin{equation}
    \begin{aligned}
        V_{\text{singlets}} (\phi_i, \Phi) & =  a_i \phi_i
        + m_{ij} \phi_i \phi_j
        + T_{ijk} \phi_i \phi_j \phi_k
        + \lambda_{ijkl} \phi_i \phi_j \phi_k \phi_l                                    \\
                                              & \quad+ T_{iHH}\phi_i (\Phi^\dagger\Phi)
        + \lambda_{ijHH}\phi_i\phi_j(\Phi^\dagger\Phi)                                  \\
    \end{aligned}
\end{equation}
with real coefficient tensors. Here, $\Phi$ describes the scalar $SU(2)_L$ doublet field of the SM and $V_\text{SM}$ denotes the scalar potential of the SM\@. An extension by complex singlets can always be
brought into this form by expanding fields and coefficients into real and
imaginary parts. Since the $\phi_i$ are pure gauge singlets they have trivial
kinetic terms that do not induce any gauge interactions, leading to the following contributions to the electroweak (EW) Lagrangian:
\begin{equation}
    \mathcal{L}_\text{EW} \supset {\left( D^\mu\Phi \right)}^\dagger D_\mu\Phi+\sum_i\partial^\mu\phi_i\partial_\mu\phi_i-V\eqdot
\end{equation}\label{eq:lag}
In addition, it is not possible to write down gauge invariant and renormalizable
interactions between a scalar singlet and any of the SM fermions. The singlets
will therefore only interact with the SM Higgs boson through the couplings of
the scalar potential and, if they acquire a non-zero vacuum expectation value
(vev), mix with the SM Higgs boson and thereby inherit some of its gauge and
Yukawa couplings.

This is also the reason why --- as long as no new interactions of the scalar
singlet fields with additional particles occur --- there is no physical
difference between a parametrization in terms of $N$ complex scalar singlet
fields or $2N$ real scalar singlet fields. Naively, one would expect that
imaginary parts of complex scalar fields are CP-odd, and mixing them with the
real parts or the SM Higgs boson would lead to CP-violation. However, due to the
singlets not having any gauge or fermion couplings it is always possible to find
a CP-transformation under which all of them are CP-even~\cite{Branco:1999fs,
Ivanov:2017dad}. Thus any pure singlet extension of the SM is a theory of only
CP-even scalars.

Singlet extensions of the SM have been subject to detailed phenomenological
studies before. This includes both extensions by a single real
singlet~\cite{Datta:1997fx,Schabinger:2005ei,Patt:2006fw,OConnell:2006rsp,Barger:2007im}
(see
Refs.~\cite{Chen:2014ask,Robens:2015gla,Robens:2016xkb,Lewis:2017dme,Ilnicka:2018def}
for recent phenomenological studies) and by a complex singlet or two real
singlets~\cite{Barger:2008jx,AlexanderNunneley:2010nw,Coimbra:2013qq,Ahriche:2013vqa,Costa:2014qga,Costa:2015llh,Ferreira:2016tcu,Chang:2016lfq,Muhlleitner:2017dkd,Dawson:2017jja}.
The models are also interesting in the context of scalar singlet dark
matter~\cite{Lerner:2009xg,Barger:2010yn,Gonderinger:2012rd,Belanger:2012zr,Ghorbani:2014gka,Jiang:2015cwa,Athron:2017kgt,Chiang:2017nmu,Athron:2018ipf,Grzadkowski:2018nbc}
and a strong first-order electroweak phase
transition~\cite{Curtin:2014jma,Jiang:2015cwa,Kotwal:2016tex,Beniwal:2017eik,Chiang:2017nmu,Cheng:2018ajh,Grzadkowski:2018nbc,Ghorbani:2019itr}.
We will focus on a specific extension of the SM by two real singlets that has
not been previously considered in the literature.

Experimentally, singlet extensions can be explored in two complementary ways at
the LHC\@. First, precise measurements of the $\SI{125}{\GeV}$ Higgs signal
rates probe the structure of the doublet-singlet mixing, as well as possible new
decay modes of the observed Higgs boson to new light scalar states. Second,
direct searches for new scalars may reveal the existence of the mostly
singlet-like Higgs bosons. For the latter the discovery prospects depend on the
singlet-doublet mixing and the new scalar's mass (both governing the production
rates), and on the decay pattern of the produced scalar state. In general,
decays directly to SM particle final states as well as to two lighter scalar
states (``Higgs-to-Higgs decays'') are possible. While some of the former decays
are already searched for by the LHC experiments, current searches for
Higgs-to-Higgs decays focus almost exclusively on the signatures $h_S\to
h_{125}h_{125}$ (where $h_S$ denotes the new Higgs state with mass above
$\SI{250}{\GeV}$)~\cite{Aaboud:2016xco,Sirunyan:2017guj,Sirunyan:2017djm,Aaboud:2018ksn,
Aaboud:2018zhh,Aaboud:2018sfw,Aaboud:2018ewm,Aaboud:2018ftw,
Sirunyan:2018two,Sirunyan:2018zkk,Sirunyan:2018iwt,Aad:2019uzh}, or $h_{125} \to
h_S h_S$ (with the $h_S$ mass below
$\SI{62.5}{\GeV}$)~\cite{Sirunyan:2018mbx,Sirunyan:2018pzn,
Aaboud:2018iil,Aaboud:2018gmx,Sirunyan:2018mot,Aaboud:2018esj,Sirunyan:2019gou}.
The model considered in the following, however, features also Higgs decays to
unidentical scalar bosons (``asymmetric decays''), Higgs decays involving only
non-SM-like Higgs bosons, as well as the possibility of successive
Higgs-to-Higgs cascade decays. All of these decay signatures have not been
experimentally explored in detail yet.\footnote{A first search result for a
symmetric Higgs-to-Higgs decay involving only non-SM Higgs states has been
presented by ATLAS in the $W^+W^-W^+W^-$ final state~\cite{Aaboud:2018ksn}.} We
will extensively discuss them in this paper and show that they lead to novel
collider signatures with sizable signal rates that are experimentally
interesting for the analysis of Run-II data as well as the upcoming LHC runs. We
provide six dedicated two-dimensional benchmark scenarios, each highlighting a
different Higgs-to-Higgs decay signature that has not been probed
experimentally. We strongly encourage the experimental collaborations to
investigate these novel signatures using current and future collider data.

This paper is structured as follows. We introduce the model in \cref{sec:model}  and summarize all relevant theoretical and experimental constraints on the parameter space in \cref{sec:TRSMconstraints}. In \cref{sec:results} we discuss the collider signatures of the model and present the impact of current LHC searches on the parameter space. In \cref{sec:benchmarks} we propose six benchmark scenarios for LHC searches for Higgs-to-Higgs decay signatures. We conclude in \cref{sec:conclusions}.

\section{The Two Real Singlet Model}\label{sec:model}

\subsection{Scalar potential and model parameters}\label{sec:TRSM_model}

The two real singlet model (TRSM) adds two real singlet degrees of freedom to
the SM\@. These are written as two real singlet fields $S$ and $X$. In order to
reduce the number of free parameters two discrete \Ztwo symmetries
\begin{equation}
    \begin{aligned}
        \Ztwo^S:\quad & S\to -S\eqcomma\ X\to X\eqcomma\ \text{SM} \to \text{SM}\eqcomma \\
        \Ztwo^X:\quad & X\to -X\eqcomma\ S \to S\eqcomma\ \text{SM} \to \text{SM}
    \end{aligned}\label{eq:Z2syms}
\end{equation}
are introduced. The most general renormalizable scalar potential invariant under
the $ \Ztwo^S\otimes\Ztwo^X$ symmetry is
\begin{equation}
    \begin{aligned}
        V & = \mu_{\Phi}^2 \Phi^\dagger \Phi + \lambda_{\Phi} {(\Phi^\dagger\Phi)}^2
        + \mu_{S}^2 S^2 + \lambda_S S^4
        + \mu_{X}^2 X^2 + \lambda_X X^4                                              \\
          & \quad+ \lambda_{\Phi S} \Phi^\dagger \Phi S^2
        + \lambda_{\Phi X} \Phi^\dagger \Phi X^2
        + \lambda_{SX} S^2 X^2\eqdot
    \end{aligned}\label{eq:TRSMpot}
\end{equation}
All coefficients in \cref{eq:TRSMpot} are real, thus the scalar potential
contains nine model parameters in total. We provide a translation of these
coefficients to the scalar potential parameters in the complex scalar singlet
parametrization in \cref{app:translation}.

Depending on the vevs acquired by the scalars different \emph{phases} of
the model can be realized. We decompose the fields (in unitary gauge) as
\begin{equation}
    \Phi = \begin{pmatrix} 0\\\frac{\phi_h + v}{\sqrt{2}}\end{pmatrix}\eqcomma\quad
    S = \frac{\phi_S + v_S}{\sqrt{2}}\eqcomma \quad
    X = \frac{\phi_X + v_X}{\sqrt{2}}
    \label{eq:fields}
\end{equation}
leading to the tadpole equations
\begin{align}
    - v \mu_{\Phi}^2 & = v^3 \lambda_\Phi + \frac{v v_S^2}{2}\lambda_{\Phi S} + \frac{v v_X^2}{2}\lambda_{\Phi X}    \\
    - v_S \mu_{S}^2  & = v_S^3 \lambda_S + \frac{v^2 v_S}{2}\lambda_{\Phi S} + \frac{v_S v_X^2}{2}\lambda_{SX}       \\
    - v_X \mu_{X}^2  & = v_X^3 \lambda_X + \frac{v^2 v_X}{2}\lambda_{\Phi X} + \frac{v_S^2 v_X}{2}\lambda_{SX}\eqdot
\end{align}
These have solutions for any values of $v$, $v_S$, $v_X$. However, to achieve
electroweak symmetry breaking $v=v_\text{SM}\approx\SI{246}{\GeV}$ is required.
If $v_S,v_X\neq0$ the \Ztwo symmetries are spontaneously broken, and the fields
$\phi_{h,S,X}$ mix into three physical scalar states. This is called the
\emph{broken phase}.

If $v_X=0$ the field $\phi_X$ does not mix with $\phi_{h,S}$, does not acquire
any couplings to SM particles, and is stabilized by the $\Ztwo^X$
symmetry.\footnote{The case of $v_S=0$ is equivalent by renaming
$S\longleftrightarrow X$.} This makes it a candidate particle for dark matter
(DM). The phenomenology of the two visible scalar states is very similar to the
real singlet extension~\cite{Chen:2014ask, Robens:2015gla, Robens:2016xkb,
Lewis:2017dme, Ilnicka:2018def}. If both singlet vevs vanish, $\phi_h$ is the SM
Higgs boson, and the two singlets both form separate dark sectors stabilized by
their respective \Ztwo symmetries. In this case collider phenomenology is (at
tree-level) only impacted by possible invisible decays of $h_{125}$ to the DM
particles.

In this work, we focus on the broken phase as it leads to the most interesting
collider phenomenology. The mass eigenstates $h_{1,2,3}$ are related to the
fields $\phi_{h,S,X}$ through the $3\times3$ orthogonal mixing matrix $R$
\begin{equation}
    \begin{pmatrix}
        h_1 \\h_2\\h_3
    \end{pmatrix} = R \begin{pmatrix}
        \phi_h \\\phi_S\\\phi_X
    \end{pmatrix}\eqdot
\end{equation}
We assume the mass eigenstates to be ordered by their masses
\begin{equation}
    M_1 \leq M_2 \leq M_3
\end{equation}
and parametrize the mixing matrix $R$ by three mixing angles $\theta_{hS}$,
$\theta_{hX}$, $\theta_{SX}$. Using the short-hand notation
\begin{equation}
    s_1\equiv\sin\theta_{hS}\eqcomma\quad s_2\equiv\sin\theta_{hX}\eqcomma\quad s_3\equiv\sin\theta_{SX}\eqcomma\quad c_1\equiv\cos\theta_{hS}\eqcomma\ \ldots
\end{equation}
it is given by
\begin{equation}
    R = \begin{pmatrix}
        c_1 c_2             & -s_1 c_2             & -s_2     \\
        s_1 c_3-c_1 s_2 s_3 & c_1 c_3+ s_1 s_2 s_3 & -c_2 s_3 \\
        c_1 s_2 c_3+s_1 s_3 & c_1 s_3-s_1 s_2 c_3  & c_2 c_3
    \end{pmatrix}\eqdot
\end{equation}
Where the angles $\theta$ can be chosen to lie in
\begin{equation}
    -\frac{\pi}{2} < \theta_{hS}, \theta_{hX}, \theta_{SX} < \frac{\pi}{2}\label{eq:angleranges}
\end{equation}
without loss of generality. In the TRSM it is possible to express the nine
parameters of the scalar potential through the three physical Higgs masses, the
three mixing angles, and the three vevs. These relations are given by
\begin{equation}
    \begin{aligned}
        \lambda_\Phi     & =\frac{1}{2 v^2} m_i^2 R_{i1}^2\eqcomma         &
        \lambda_S        & =\frac{1}{2 v_S^2}  m_i^2 R_{i2}^2\eqcomma      &
        \lambda_X        & =\frac{1}{2 v_X^2}  m_i^2 R_{i3}^2\eqcomma        \\
        \lambda_{\Phi S} & =\frac{1}{v v_S}  m_i^2 R_{i1} R_{i2}\eqcomma   &
        \lambda_{\Phi X} & =\frac{1}{v v_X}  m_i^2 R_{i1} R_{i3}\eqcomma   &
        \lambda_{S X}    & =\frac{1}{v_S v_X}  m_i^2 R_{i2} R_{i3}\eqcomma
    \end{aligned}\label{eq:TRSMlams}
\end{equation}
where a sum over $i$ is implied. Fixing one of the Higgs masses  to the mass of
the observed Higgs boson, $M_a \simeq \SI{125}{\GeV}$, and fixing the Higgs
doublet vev to its SM value, $v \simeq \SI{246}{\GeV}$, leaves seven free input
parameters of the TRSM\@:
\begin{equation}
    M_b\eqcomma\ M_c\eqcomma\ \theta_{hS}\eqcomma\ \theta_{hX}\eqcomma\ \theta_{SX}\eqcomma\ v_S\eqcomma\ v_X\eqcomma\label{eq:TRSMpars}
\end{equation}
{with $a\neq{}b\neq{}c\in\lbrace1,2,3\rbrace$.}

This is an important practical difference between the TRSM and another
well-studied extension of the SM with two real singlet degrees of freedom, the
CxSM~\cite{Coimbra:2013qq,Costa:2015llh}. The CxSM is expressed in terms of a
complex singlet with a softly broken $U(1)$ symmetry of the singlet phase
imposed on the scalar potential. This more stringent symmetry leaves only seven
model parameters such that one of the physical scalar masses and one of the
singlet vevs are dependent parameters. In contrast, the TRSM is consistent for
\emph{any} combination of masses, mixing angles, and vevs, and therefore allows
to cover the full possible kinematic phase space of Higgs-to-Higgs decay
signatures, as we will do when defining the benchmark scenarios in
\cref{sec:benchmarks}.

As in all pure singlet extensions the couplings of the scalar boson $h_a$ ($a\in \{ 1,2,3 \} $) to
all SM particles are given by the SM prediction rescaled by a common factor
\begin{equation}
    \kappa_a = R_{a1},
\end{equation}
{where $R_{a1}$ denotes} the doublet admixture of the mass eigenstate $h_a$. Due to the orthogonality
of the mixing matrix these obey the important sum rule
\begin{equation}
    \sum_{a=1}^3 \kappa_a^2 = 1\eqdot\label{eq:TRSMsumrule}
\end{equation}

\subsection{Collider Phenomenology}\label{sec:TRSMcollpheno}

The triple Higgs couplings are of special phenomenological interest in the
TRSM\@. Using \cref{eq:TRSMlams} they can be expressed directly through the
input parameters of \cref{eq:TRSMpars}. The coupling $\tilde\lambda_{abb}$ of
$h_a h_b h_b$ is defined through
\begin{equation}
    V \supset \frac{h_a h_b^2}{2} \left( \sum_j \frac{R_{a j}\,R^2_{b j}}{v_j}\right) \left( M_a^2+2 M_b^2 \right)
    \equiv \frac{1}{2} \tilde{\lambda}_{abb}h_a h_b^2\eqdot\label{eq:symmetriccoup}
\end{equation}
Similarly, the coupling of three different scalars is given by
\begin{equation}
    V\supset h_a h_b h_c \left( \sum_j \frac{R_{a j} R_{b j} R_{c j}}{v_j}\right) \left( \sum_i  M_i^2 \right)
    \equiv \tilde{\lambda}_{abc}h_a h_b h_c \eqcomma\label{eq:asymmetriccoup}
\end{equation}
and the triple Higgs self-coupling $\tilde\lambda_{aaa}$ reads
\begin{equation}
    V\supset \frac{h^3_a}{2} \left( \sum_j \frac{R^3_{a j}}{v_j}\right) M_a^2
    \equiv \frac{1}{3!}\tilde{\lambda}_{aaa} h_a^3\eqdot
\end{equation}
With these definitions the tree-level partial decay width of a scalar $h_a$ into two scalars $h_b$ and
$h_c$ (where $b=c$ is allowed) is then given by
\begin{equation}
    \Gamma_{a \rightarrow bc} = \frac{\tilde{\lambda}_{abc}^2}{16 \pi M_a^3} \sqrt{\lambda({M_a^2}, M_b^2, M_c^2)} \frac{1}{1+\delta_{bc}} \Theta(M_a-M_b-M_c)\eqcomma\label{eq:higgstohiggs}
\end{equation}
with
\begin{equation}
    \lambda(x_1,x_2,x_3) \equiv \sum_{i}x^2_i-\sum_{i,j \neq i} x_i x_j\eqdot
\end{equation}

With this information, the phenomenology of a TRSM Higgs boson $h_a$ can be
fully obtained from the predictions for a SM-like Higgs boson $h_\text{SM}$ of
the same mass. Throughout this work we employ the narrow width approximation to
factorize  production cross sections and branching ratios (BRs).

For a certain production process (e.g.~gluon gluon fusion) the cross
section, $\sigma$, for $h_a$ with mass $M_a$ can be obtained from the
corresponding SM Higgs production cross section, $\sigma_\text{SM}$, by
simply rescaling
\begin{equation}
    \sigma(M_a) = \kappa_a^2 \cdot \sigma_\text{SM} ( M_a)\eqdot\label{eq:cxnscaling}
\end{equation}
Since $\kappa_a$ rescales all Higgs couplings to SM particles,
\cref{eq:cxnscaling} is exact up to genuine EW corrections involving Higgs
self-interactions. In particular, this holds to all orders in QC\@D.

The scaling factor $\kappa_a$ also rescales universally the partial widths of
$h_a$ decays into SM particles, which in turn leads to a rescaling of the SM
total width as
\begin{equation}
    \Gamma(h_a\to\text{SM}; M_a) = \kappa_a^2 \cdot \Gamma_\text{tot}(h_\text{SM}; M_a),\label{eq:widthscaling}
\end{equation}
where $\Gamma( h_a \to \text{SM}; M_a)$ denotes the sum of all partial widths of $h_a$ into SM particle final states.
Note that this alone can never change the BR predictions of $h_a$ into SM particles. Using the
results of \cref{eq:higgstohiggs} we can obtain the BRs of $h_a$ decays to other scalar bosons, $h_a \to h_b h_c$:
\begin{equation}
    \text{BR}(h_a\to h_b h_c) = \frac{\Gamma_{a\to bc}}{\kappa_a^2~\Gamma_\text{tot}(h_\text{SM}) + \sum_{xy} \Gamma_{a\to xy}}\eqdot
\end{equation}
Denoting the sum of these ``new physics'' (NP) decay rates to scalar boson final states as
\begin{equation}
    \text{BR}(h_a\to\text{NP}) \equiv \sum_{b,c} \text{BR}(h_a\to h_b h_c)\eqcomma
\end{equation}
the BRs of $h_a$ decays into any final state $F_\text{SM}$ composed entirely of SM fermions and gauge bosons are given by
\begin{equation}
    \text{BR}(h_a\to F_\text{SM}) = \left(1 - \text{BR}(h_a\to\text{NP})\vphantom{x^2}\right) \text{BR}(h_\text{SM}\to F_\text{SM}) \eqdot
    \label{eq:BRSM}
\end{equation}
One important special case is that in the absence of BSM decay modes --- which
is always the case for the lightest Higgs bosons $h_1$ --- $h_a$ has BRs
identical to a SM-like Higgs boson of the same mass.

\begin{figure}
    \centering
    \subfloat[low mass]{\includegraphics[width=0.49\textwidth]{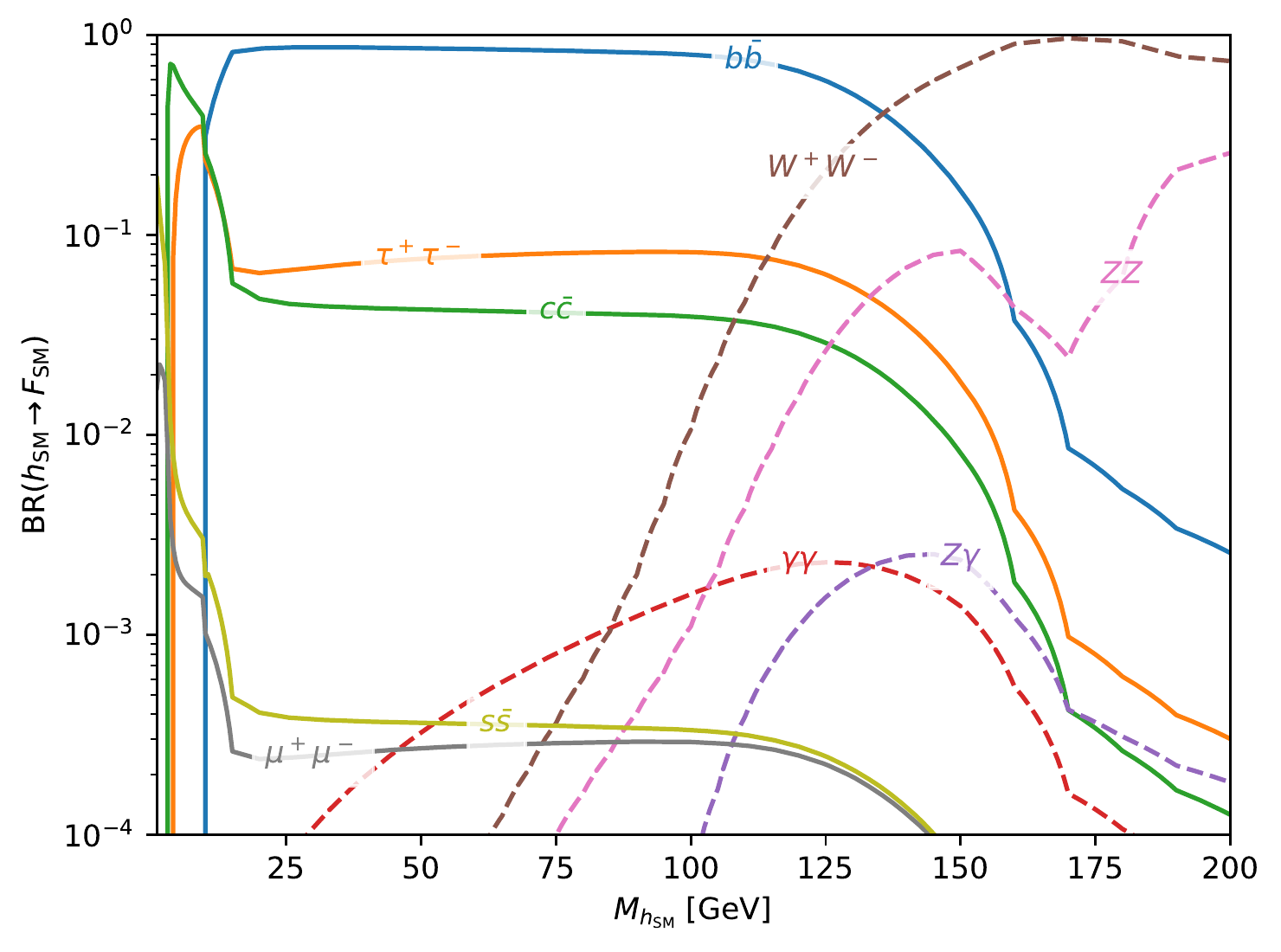}}
    \subfloat[high mass]{\includegraphics[width=0.49\textwidth]{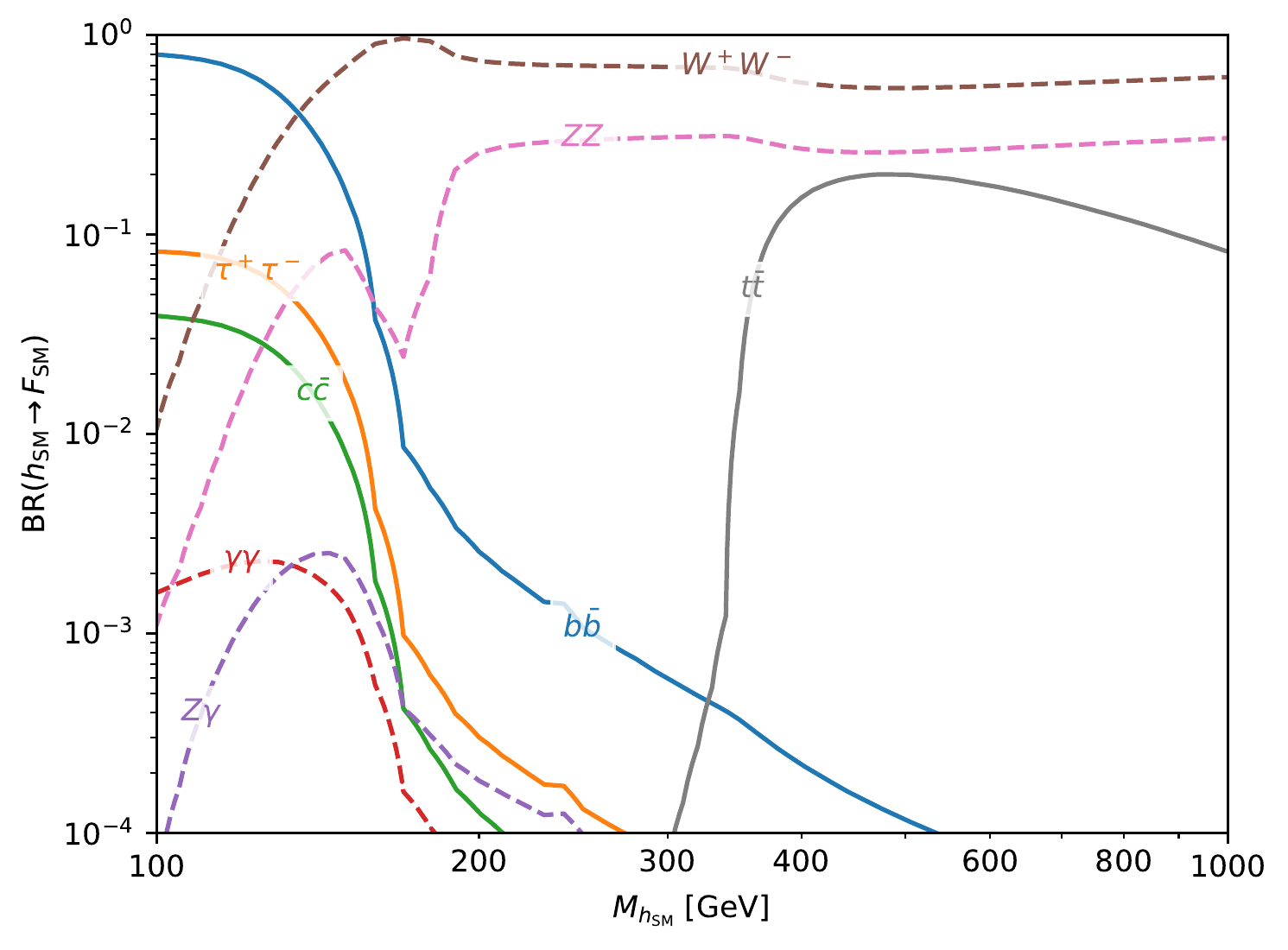}}
    \caption{Decay branching ratios of a SM-like Higgs boson, $h_\text{SM}$, for
            various SM particle final states, $F_\text{SM}$, as a function of
            its mass, $M_{h_\text{SM}}$, in the mass range from $\SI{1}{\GeV}$
            to $\SI{1}{\TeV}$, split into the low mass region (\emph{left
            panel}) and the high mass region (\emph{right panel}). The numerical
            values are taken from Ref.~\cite{deFlorian:2016spz}, see text for
            further details.}\label{fig:SMBRs}
\end{figure}

\Cref{fig:SMBRs} shows the decay branching ratios of a SM-like Higgs boson
$h_\text{SM}$ as a function of its mass. As long as $\BR{h_a\to\text{NP}}=0$,
\ie if no Higgs-to-Higgs decays are possible for $h_a$, the scalar boson $h_a$
has exactly the BRs shown in \cref{fig:SMBRs}. The numerical values are taken
from Ref.~\cite{deFlorian:2016spz}, based on state-of-the-art evaluations using
\texttt{HDECAY}~\cite{Djouadi:1997yw,Butterworth:2010ym,Djouadi:2018xqq}
and
\texttt{Profecy4F}~\cite{Bredenstein:2006rh,Bredenstein:2006nk,Bredenstein:2006ha}.

\section{Setup of the Parameter Scan}\label{sec:TRSMconstraints}

In order to assess the phenomenologically viable regions of the parameter space
we apply all relevant theoretical and experimental constraints, which are
discussed in the following. We furthermore provide details of our numerical
setup.

\subsection{Theoretical Constraints}

Unitarity constraints provide important upper bounds on the multi-scalar
couplings and the scalar masses. In the TRSM we have derived perturbative
unitarity bounds in the high energy limit by requiring the eigenvalues $M^i$ of
the 2-to-2 scalar scattering matrix $M$ to fulfill
\begin{equation}
    |M^i| < 8\pi\eqdot
\end{equation}
The resulting constraints on the parameters of the scalar potential are
\begin{align}
    \left|\lambda_\Phi\right|                                                                 & < 4\pi ,  \\
    \left|\lambda_{\Phi S}\right|,\,\left|\lambda_{\Phi X}\right|,\,\left|\lambda_{SX}\right| & < 8\pi ,  \\
    |a_1|,\, |a_2|,\, |a_3|                                                                   & < 16\pi ,
\end{align}
where $a_{1,2,3}$ are the three real roots of the cubic polynomial
\begin{equation}
    \begin{aligned}
        P(x) & \equiv  x^3+x^2 (-12 \lambda_\Phi-6 \lambda_S-6 \lambda_X)                                            \\
             & \quad +x \left(72 \lambda_\Phi (\lambda_S+\lambda_X) -4 (\lambda_{\Phi S}^2+ \lambda_{\Phi X}^2)+36
        \lambda_S \lambda_X-\lambda_{SX}^2\right)                                                                    \\
             & \quad +12 \lambda_\Phi \lambda_{SX}^2+24 \lambda_{\Phi S}^2 \lambda_X+24 \lambda_{\Phi X}^2 \lambda_S
        -8 \lambda_{\Phi S} \lambda_{\Phi X} \lambda_{SX}-432 \lambda_\Phi \lambda_S \lambda_X \eqdot
    \end{aligned}
\end{equation}

Closed form conditions for boundedness of the scalar potential, \cref{eq:TRSMpot},
have been derived in~\cite{Kannike:2012pe,Kannike:2016fmd}. In our notation they read
\begin{equation}
    \begin{aligned}
        \lambda_\Phi,\lambda_S,\lambda_X                                                                                               & > 0\eqcomma  \\
        \overline{\lambda}_{\Phi S} \equiv \lambda_{\Phi S} + 2 \sqrt{\lambda_\Phi\lambda_S}                                           & > 0 \eqcomma \\
        \overline{\lambda}_{\Phi X} \equiv \lambda_{\Phi X} + 2 \sqrt{\lambda_\Phi\lambda_X}                                           & > 0 \eqcomma \\
        \overline{\lambda}_{SX} \equiv \lambda_{SX} + 2 \sqrt{\lambda_S\lambda_X}                                                      & > 0 \eqcomma \\
        \sqrt{\lambda_S}\lambda_{\Phi X}
        + \sqrt{\lambda_X}\lambda_{\Phi S}
        + \sqrt{\lambda_\Phi}\lambda_{SX}
        + \sqrt{\lambda_\Phi\lambda_S\lambda_X} + \sqrt{\overline{\lambda}_{\Phi S}\overline{\lambda}_{\Phi X}\overline{\lambda}_{SX}} & > 0\eqdot
    \end{aligned}
\end{equation}

It has been proven in Ref.~\cite{Ferreira:2016tcu} that at tree-level a vacuum of the form of \cref{eq:fields} is always the global minimum of the scalar TRSM potential. Therefore no additional constraints from vacuum decay need to be considered.
\subsection{Experimental Constraints}
\label{sec:expconstraints}
We use the oblique parameters $S$, $T$ and $U$~\cite{Altarelli:1990zd,Peskin:1990zt,Peskin:1991sw,Maksymyk:1993zm} to parametrize constraints from
electroweak precision measurements, which are compared to the latest fit results~\cite{Haller:2018nnx}. The results of
Refs.~\cite{Grimus:2007if,Grimus:2008nb} are applicable to the TRSM to obtain
model predictions for $S$, $T$ and $U$.\footnote{The $W$-boson mass could be
used as a single precision observable for models with new particle content, see
e.g.~Ref.~\cite{Lopez-Val:2014jva} for a discussion within the real singlet
extension.  We checked the TRSM with an extension of the code presented in
Ref.~\cite{Lopez-Val:2014jva} and compared to the updated experimental value
$M_W = \SI{80.379 \pm 0.012}{\GeV}$~\cite{Schael:2013ita,Aaltonen:2013vwa,Aaboud:2017svj,Tanabashi:2018oca}. We
found no relevant additional constraints from $M_W$ in this model.} Flavor
constraints are not relevant as the singlets do not change the Yukawa sector. We
use \HiggsBounds-5.4.0~\cite{Bechtle:2008jh, Bechtle:2011sb, Bechtle:2013gu,
Bechtle:2013wla, Bechtle:2015pma, HBwebsite} to check for agreement with the
bounds from searches for additional Higgs bosons.

Important constraints on the model parameter space arise from the signal rate
measurements of the observed $\SI{125}{\GeV}$ Higgs boson, which we denote by
$h_{125}$ in the following. These constraints are especially relevant in singlet
extensions as there are effectively only two BSM parameters that enter the
phenomenology of the $h_{125}$: its coupling scale factor $\kappa_{125}$ and its
decay rate $\text{BR}(h_{125}\to\text{NP})$ into new particles (see
\cref{sec:TRSMcollpheno}).

We use \HiggsSignals-2.3.0~\cite{Stal:2013hwa, Bechtle:2013xfa,
Bechtle:2014ewa, HSwebsite} to test for agreement with the observations at the
$2\sigma$ level using a profiled likelihood ratio test with the SM as
alternative hypothesis. In practice, the likelihood ratio test statistic is
calculated via the difference between the log-likelihoods, which in turn is
approximated as $\Delta \chi^2 = \chi^2 - \chi^2_\text{SM}$ within
\HiggsSignals. As we have two relevant statistical degrees of freedom that can
influence the Higgs signal rate predictions (see above), we obtain the $2\sigma$
confidence region by demanding $\Delta \chi^2 \le 6.18$. \HiggsSignals-2.3.0
contains the latest measurements from ATLAS~\cite{ATLAS:2019slw} and
CMS~\cite{Sirunyan:2017elk,Sirunyan:2017dgc,Sirunyan:2018egh,CMS:1900lgv,Sirunyan:2018hbu,CMS-PAS-HIG-18-030,Sirunyan:2018shy,CMS:2018dmv,CMS:2019chr,CMS:2019pyn}
with up to $\sim \SI{137}{\fb^{-1}}$ of data collected during Run-II at a
center-of-mass energy of $\SI{13}{\TeV}$, as well as the ATLAS and CMS combined
measurements from Run-I~\cite{Khachatryan:2016vau}.

A further complication may arise in this model in case that two or even all
three scalar bosons have a mass around $\SI{125}{\GeV}$. \HiggsSignals then
automatically takes into account a possible superposition of their signals in
the test against the Higgs rate measurements {by incoherently summing the
contributions of all scalars. This approach neglects any possible interference
effects}, see Ref.~\cite{Bechtle:2013xfa} for details. A similar approach is
employed in \HiggsBounds to combine multiple scalars that lie within the
experimental mass resolution.

\begin{figure}
    \centering
    \includegraphics[width=0.55\textwidth, trim=0cm 1cm 0cm 1cm]{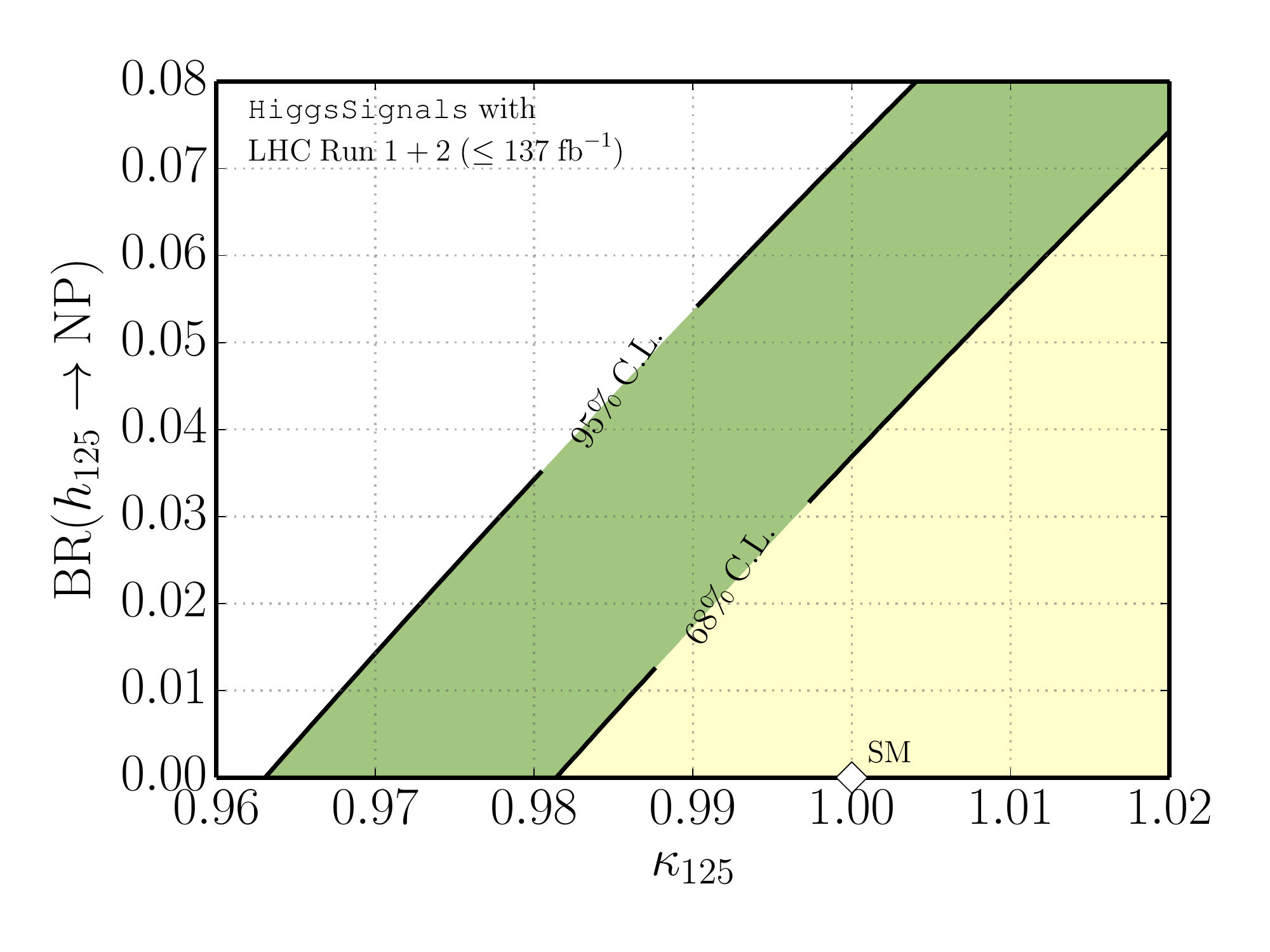}
    \caption{Constraints from Higgs signal rate measurements on the parameters
        $\kappa_{125}$ and $\text{BR}(h_{125}\to\text{NP})$ as obtained from
        \HiggsSignals-2.3.0.}\label{fig:kappaBRNP}
\end{figure}

Assuming only one scalar boson is responsible for the observed signal at
$\SI{125}{\GeV}$, we show the constraints from Higgs signal rate measurements in
the (simplified) two-dimensional parameter plane ($\kappa_{125}$,
$\text{BR}(h_{125}\to\text{NP})$) in \cref{fig:kappaBRNP}.\footnote{The expected
sensitivity of Higgs rate measurements at the high-luminosity (HL)-LHC in this
parameter plane has been presented in Section~6 of Ref.~\cite{Cepeda:2019klc}.}
If no BSM decay modes of $h_{125}$ exist, a lower bound on $\kappa_{125}>0.963$
at $\SI{95}{\%}$ C.L.\ is obtained. For the other limiting case of exactly
SM-like couplings, $\kappa_{125}=1$, we find a limit of
$\BR{h_{125}\to\text{NP}}<\SI{7.3}{\%}$. The $2\sigma$ limit between these two
limiting cases follows approximately a linear slope. The region $\kappa>1$ is
only included for completeness in \cref{fig:kappaBRNP} but cannot be realized in
the TRSM, see \cref{eq:TRSMsumrule}. Note that this analysis is applicable
to any model where a singlet scalar mixes with the Higgs boson. This bound can
\eg be applied to Higgs portal models, where it gives a stronger constraint than
direct measurements of $\BR{h_{125}\to
\text{invisible}}$~\cite{Sirunyan:2018owy,Aaboud:2019rtt}.

\subsection{Numerical Evaluation}\label{sec:TRSMscan}
Based on these constraints we performed a large scan of the TRSM parameter
space using an updated private version of the code
\texttt{ScannerS}~\cite{Coimbra:2013qq,Ferreira:2014dya,Costa:2015llh,Muhlleitner:2016mzt}.
For the determination of viable regions in the parameter space, we apply all of the constraints described above. Note that bounds from signal strength measurements are evaluated with \HiggsSignals for each point individually. This guarantees that
the possibility that two or even all three Higgs bosons may have masses close to $\SI{125}{\GeV}$ and therefore contribute to the observed signal is correctly accounted for.

We parametrize the model via the input parameters given in \cref{eq:TRSMpars}.
For the numerical results presented in \cref{sec:results} we independently
sample from uniform distributions for each parameter. We allow for the
non-$h_{125}$ Higgs masses and the singlet vevs to lie within
\begin{equation}
    \SI{1}{\GeV} \le M_b, M_c, v_X, v_S \le \SI{1}{\TeV}
\end{equation}
and vary the mixing angles throughout their allowed range,
\cref{eq:angleranges}. We only keep parameter points that pass all constraints.
For the benchmark scenarios in \cref{sec:benchmarks}, we fix all parameters
apart from the non-SM scalar masses, and scan the two-dimensional parameter
space in a grid within the defined parameter ranges.

{The singlet vevs are only mildly constrained by current experimental results
   while theoretical constraints --- in particular perturbative unitarity ---
   only require them to not be substantially smaller than the scalar masses. On the other hand, as they enter the triple scalar couplings,
\cref{eq:symmetriccoup,eq:asymmetriccoup}, they can influence the relative importance of different
Higgs-to-Higgs decay modes without changing the remaining phenomenology. We therefore expect that future results from LHC searches for Higgs-to-Higgs decays will be able to constrain the vacuum structure of the singlet fields.}

For the SM Higgs production cross sections and decay rates we use the
predictions from Refs.~\cite{Heinemeyer:2013tqa,deFlorian:2016spz}. The
$h_\text{SM}$ production cross sections and total width are rescaled according
to \cref{eq:widthscaling,eq:cxnscaling} and combined with leading-order decay
widths for the Higgs-to-Higgs decays from \cref{eq:higgstohiggs}. For the
$h_{125}$ production rates, we use the results of the N$^{3}$LO calculation in
the gluon gluon fusion (ggF) channel~\cite{Anastasiou:2016cez}. This calculation
uses an effective description of the top-induced contributions. For scalar
bosons with masses $M_a\neq\SI{125}{\GeV}$ we instead employ results from the
NNLO+NNLL calculation~\cite{Heinemeyer:2013tqa} that accounts for top-quark mass
induced effects up to NLO\@. Indeed, we find that the predictions of these
calculations differ sizably for scalar boson masses $M_a\gtrsim2m_t$, for
instance, at $M_a = \SI{400}{\GeV}$,
\begin{equation}
    \left.\frac{\sigma_\text{NNLO+NNLL}}{\sigma_\text{N3LO}}\right|_{M=\SI{400}{\GeV}}\sim 3\eqdot
\end{equation}

In the following discussion of collider signatures we assume the production of a
single scalar state via the dominant ggF process. In some cases, though, it
might be worthwhile to investigate the discovery potential of the subdominant
Higgs production processes of vector boson fusion or Higgs-Strahlung, $pp \to
V\phi$~($V = W^\pm, Z$), as these give additional trigger options and may help
to reduce the background. We leave the detailed exploration of the prospects for
various production modes to future studies.

Scalar pair production can proceed through a top-quark box diagram in addition
to single Higgs production followed by a Higgs-to-Higgs decay. For pair
production of $h_{125}$ these diagrams and their interference effects with the
resonant production are known to be important (see e.g.
Refs.~\cite{Dawson:2015haa,Carena:2018vpt,DiMicco:2019ngk,Basler:2019nas}).  For
cases other than $h_{125}$-pair production the box diagrams are significantly
less important as they are always suppressed by the small $\kappa$ factors of
the non-$h_{125}$ scalars. Signal-signal interference effects between different
resonant scalars of similar mass have also been shown to significantly impact
di-Higgs production cross sections~\cite{Basler:2019nas}. {However, such mass
configurations play no important role for most of the scenarios in the following
discussion.}

\section{Implications of current collider searches}
\label{sec:results}

\subsection{Signatures of New Scalars decaying to SM particles}
\begin{figure}
    \centering
    \subfloat{\includegraphics[width=.49\textwidth]{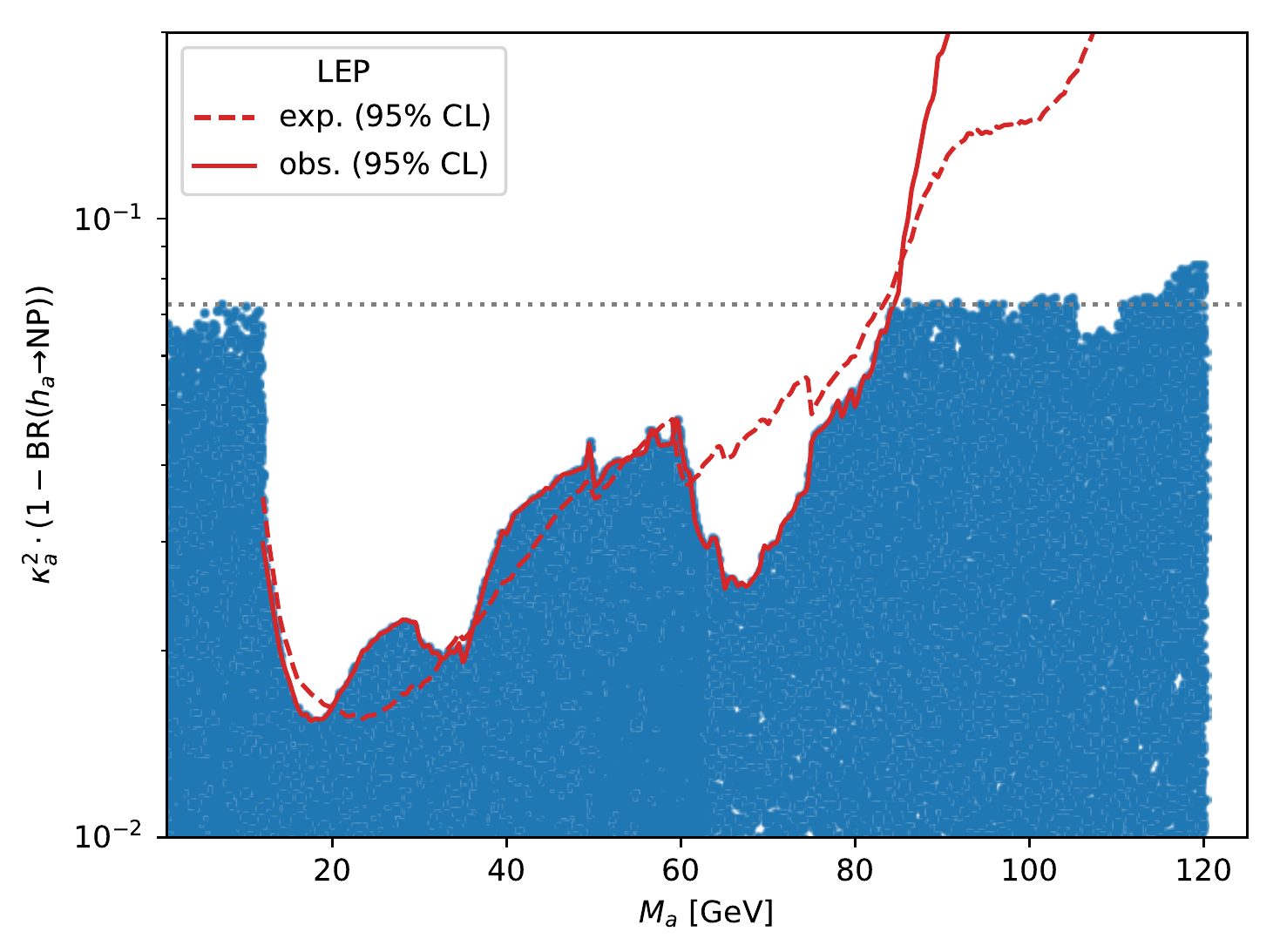}}
    \subfloat{\includegraphics[width=.49\textwidth]{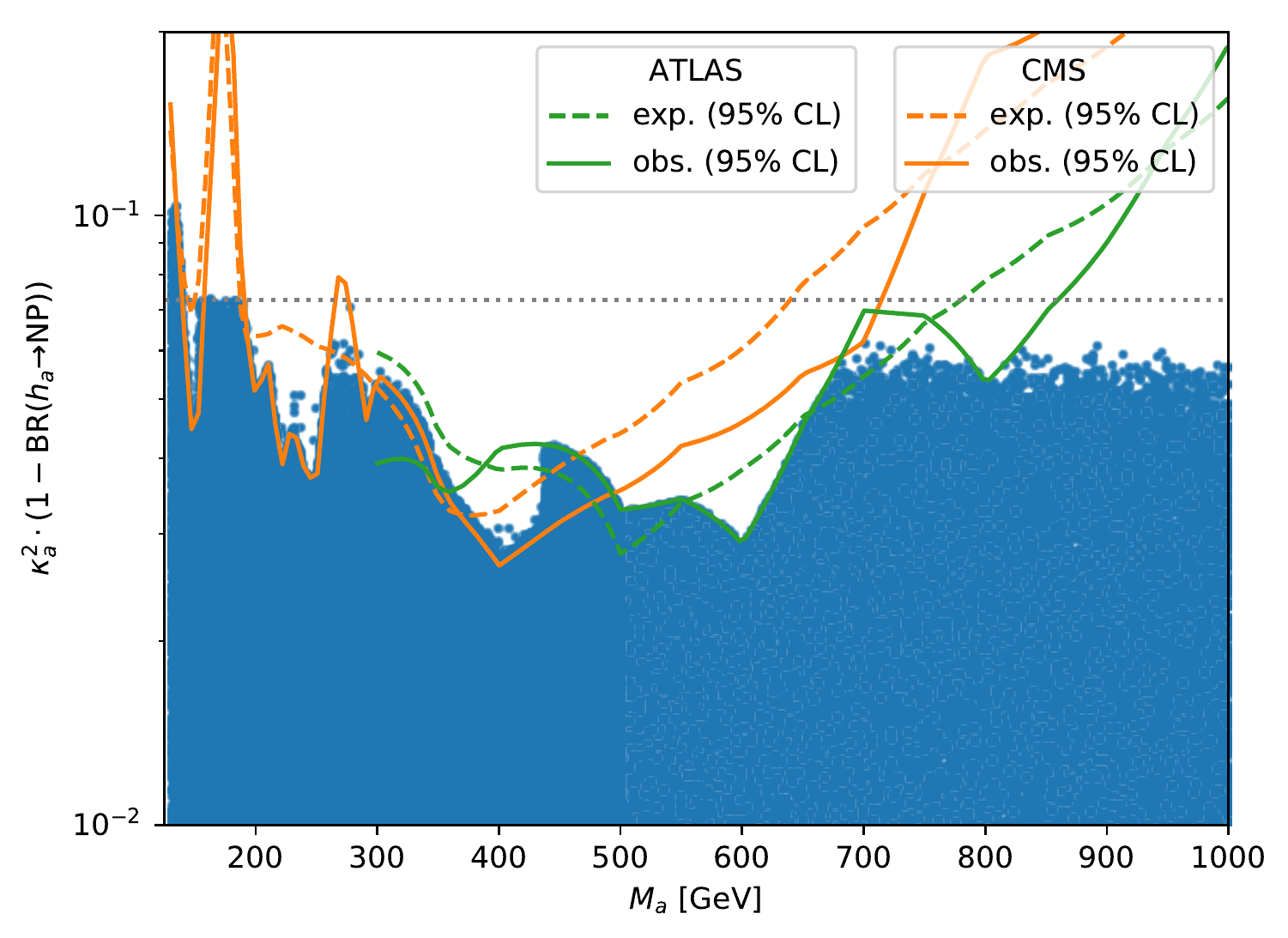}}
    \caption{SM-normalized signal rate for additional Higgs bosons decaying to
        SM particle final states as a function of its mass, $M_a$, for all parameter
        points passing all relevant constraints (\emph{blue points}). In the low
        mass region (\emph{left panel}) we include the observed (\emph{solid line})
        and expected (\emph{dashed line}) limit from LEP searches in the $ee\to h_a
            Z \to bb Z$ channel~\cite{Schael:2006cr}. In the high mass region
        (\emph{right panel}) the ATLAS~\cite{Aaboud:2018bun} and
        CMS~\cite{Sirunyan:2018qlb} observed and expected limits from the latest $p
            p \to h_a \to ZZ/WW$ searches are displayed. The \emph{dotted gray line}
        indicates the indirect limit on $\kappa^2$ from Higgs rate measurements.}\label{fig:directsearches}
\end{figure}

The additional scalar bosons $h_a$ ($h_a \neq h_{125}$) can decay directly to SM particles. The branching ratios of the various SM particle final states ($F_\text{SM}$) are obtained according to \cref{eq:BRSM}, and their relative rates (i.e.~the ratios of branching ratios for different $F_\text{SM}$ decay modes) are identical to the corresponding SM predictions for a Higgs boson with mass $M_a$. The rate for $h_a$ signal processes leading to $F_\text{SM}$, normalized to the corresponding SM prediction, can therefore be expressed as
\begin{equation}
    \frac{\sigma(pp\to h_a (+X)) \times \mathrm{BR}(h_a \to F_\text{SM})}{\sigma_\text{SM}(pp\to h_\text{SM} (+X)) \times \mathrm{BR}(h_\text{SM} \to F_\text{SM})}  = \kappa_a^2 \cdot \left(1   - \mathrm{BR}(h_a \to \text{NP})\right) \eqdot
\end{equation}
This quantity is shown in \cref{fig:directsearches} as a function of $M_a$ in the low mass region (\emph{left panel}) and high mass region (\emph{right panel}) for the sampled parameter points that pass all relevant constraints (see \cref{sec:TRSMconstraints}). For $M_a$ roughly between $12$ and $\SI{85}{\GeV}$ LEP searches for $e^+e^- \to h_a Z \to b\bar{b} Z$~\cite{Schael:2006cr} lead to an upper limit on the possible signal rate, as shown by the red lines in \cref{fig:directsearches}~(\emph{left}). At larger mass values $\gtrsim \SI{190}{\GeV}$, the upper limit originates from LHC searches for $pp \to h_a \to W^+W^-$ and $ZZ$. The latest ATLAS~\cite{Aaboud:2018bun} and CMS~\cite{Sirunyan:2018qlb} limits are overlaid as green and orange lines, respectively, in \cref{fig:directsearches}~(\emph{right}). For very large mass values $\gtrsim \SI{700}{\GeV}$ direct LHC searches are not yet sensitive to probe the parameter space. In addition, we include in \cref{fig:directsearches} the upper limit inferred indirectly via the sum rule, \cref{eq:TRSMsumrule}, from the rate measurements of the $\SI{125}{\GeV}$ Higgs state. These lead to an upper limit of $\kappa_a^2 \le 7.3\%$ (see \cref{sec:expconstraints}), except in the mass region around $\SI{125}{\GeV}$ where $h_a$ potentially contributes to the observed Higgs signal.

\subsection{Signatures of Resonant Scalar Pair Production}
The model allows for resonant scalar pair-production at the LHC, or, in other
words, the direct production of a single scalar $h_a$ followed by the
``symmetric'' or ``asymmetric'' decay into identical or different scalar states,
respectively. Specifically,
\begin{align}
    pp\rightarrow h_a~(+ X)\rightarrow h_b h_b~(+ X),  \label{eq:process_sym} \\
    pp\rightarrow h_3~(+ X)\rightarrow h_1 h_2~(+ X), \label{eq:process_asym}
\end{align}
where, in the symmetric case, \cref{eq:process_sym}, $a=2$, $b=1$ or $a=3$,
$b\in \{1,2\}$, and $X$ denotes not further defined objects that may be produced
in association with the scalar state (e.g., jets, vector bosons, etc.). The
$h_{125}$ can be either of the three scalar states $h_a$ ($a\in \{1,2,3\}$).

\begin{figure}
    \centering
    \subfloat{\includegraphics[width=0.49\textwidth]{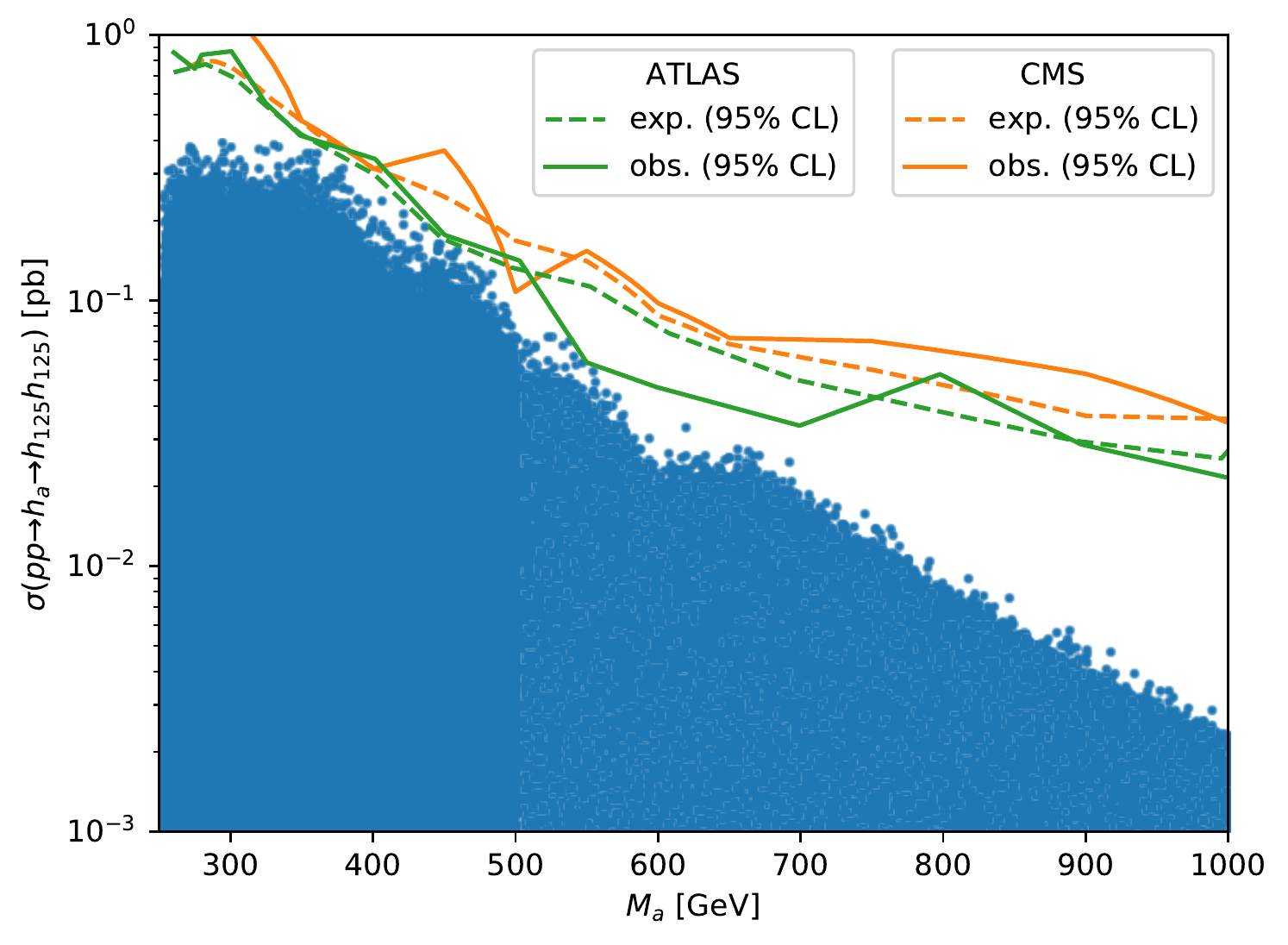}}
    \subfloat{\includegraphics[width=.49\textwidth]{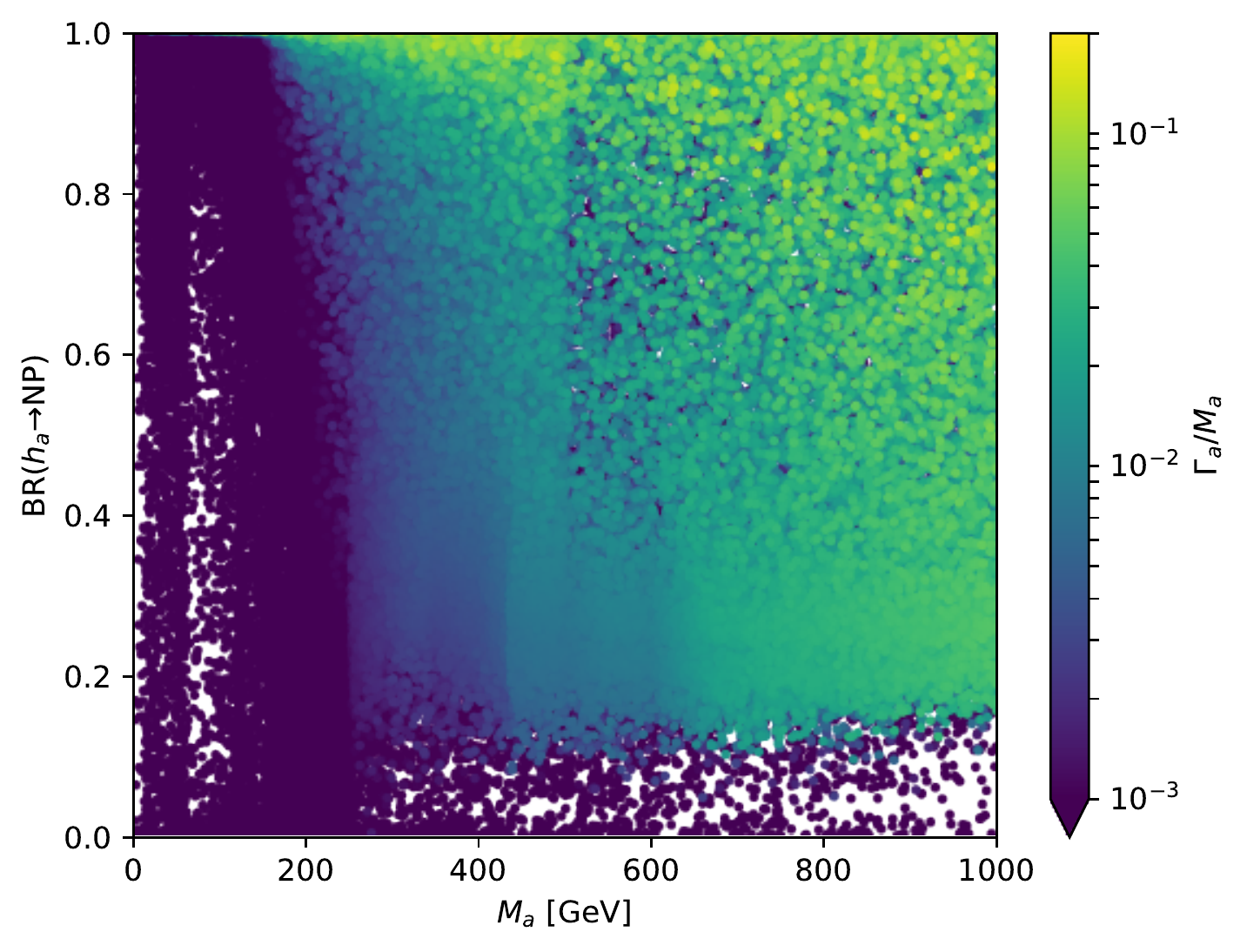}}
    \caption{\emph{Left panel:} Signal rate for the process $pp \to h_a \to h_{125} h_{125}$ at the $\SI{13}{\TeV}$ LHC as a function of the $h_a$ mass, $M_a$, for all parameter points passing all relevant constraints (\emph{blue points}). The current expected and observed upper limits on this process from ATLAS~\cite{Aad:2019uzh} (\emph{green lines}) and CMS~\cite{Sirunyan:2018two} (\emph{orange lines}) are overlaid. \emph{Right panel:} Total decay width over mass, $\Gamma_a / M_a$, of the resonant scalar as a function of $M_a$ and the decay rate $\mathrm{BR}(h_a \to \text{NP})$. Parameter points with larger $\Gamma_a / M_a$ values are plotted on top of points with smaller values.}
    \label{fig:h125pairs}
\end{figure}

Processes of the symmetric type, \cref{eq:process_sym}, leading to pair
production of $h_{125}$ are already being investigated, see \eg
Refs.~\cite{Aaboud:2016xco,Sirunyan:2017guj,Sirunyan:2017djm,
Aaboud:2018zhh,Aaboud:2018ksn,Aaboud:2018sfw,Aaboud:2018ewm,Aaboud:2018ftw,
Sirunyan:2018two,Sirunyan:2018zkk,Sirunyan:2018iwt,Aad:2019uzh} for recent
LHC~Run-II searches. \Cref{fig:h125pairs}~(\emph{left}) shows the
$\SI{13}{\TeV}$ LHC signal rate for the resonant scalar pair production process
$pp\to h_a \to h_{125}h_{125}$ ($a\in \{ 2,3\}$) as a function of the $h_a$
mass, $M_a$. \Cref{fig:h125pairs} contains the complete sample of allowed
parameter points generated according to \cref{sec:TRSMscan}. Overlaid are the
most recent experimental limits on this process from the
ATLAS~\cite{Aad:2019uzh} and CMS~\cite{Sirunyan:2018two} collaborations.
\Cref{fig:h125pairs}~(\emph{left}) illustrates that experimental searches in
this channel are beginning to directly constrain the TRSM for resonance masses
between around $\SI{380}{\GeV}$ and $\SI{550}{\GeV}$. In contrast, LHC
searches~\cite{Sirunyan:2018mbx,Sirunyan:2018pzn,
Aaboud:2018iil,Aaboud:2018gmx,Sirunyan:2018mot,Aaboud:2018esj,Sirunyan:2019gou}
for the inverted signature of single-production of $h_{125}$ which then decays
into a pair of light $h_a$ ($a\in \{1,2\})$ are not yet sensitive, as the
indirect constraints from Higgs signal rates on the possible new decay modes,
$\text{BR}(h_{125} \to \text{NP}) \le 7.3\%$ (see \cref{fig:kappaBRNP}), are
much stronger than the direct limits from these searches.\footnote{Currently,
the strongest limit from $h_{125} \to h_a h_a$ searches is obtained in the
$b\bar{b}\tau^+\tau^-$ final state~\cite{Sirunyan:2018pzn} at $M_a \simeq
\SI{35}{\GeV}$, amounting to around $\mathrm{BR}(h_{125}\to h_a h_a) \le
\SI{25}{\%}$ (assuming $h_a$ to decay exclusively to SM particles) in the TRSM.}
Both of these processes are under active experimental investigation and we
expect the bounds to improve in the future.

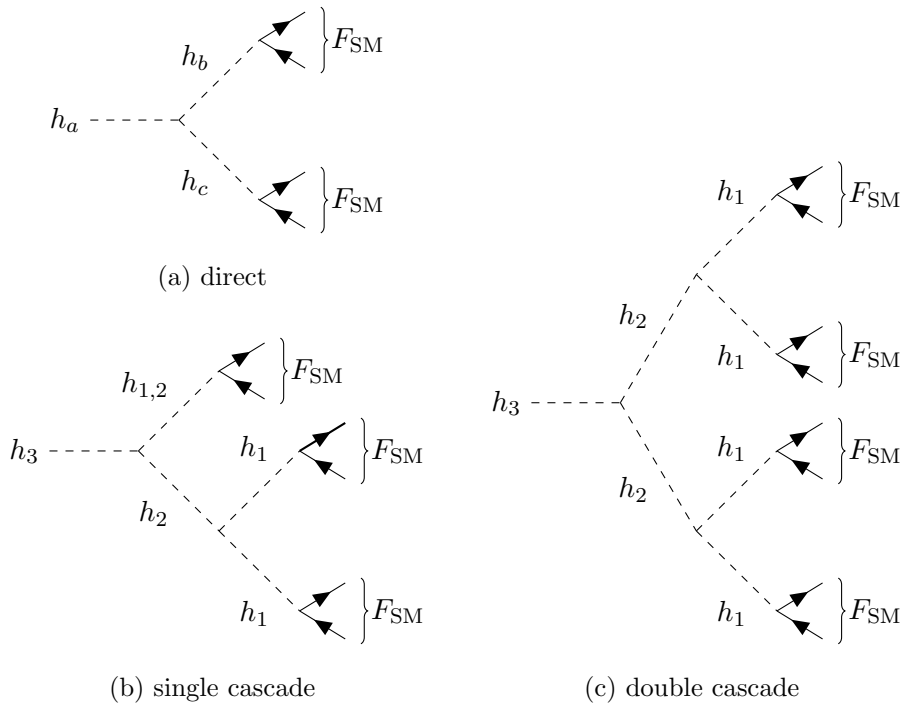
\begin{figure}
    \centering
    \begin{minipage}[b]{0.4\linewidth}
        \subfloat[direct\label{sfig:direct}]{
            \begin{tikzpicture}
                \begin{feynman}
                    \vertex (i1) {\(h_{a} \)};
                    \vertex [right=of i1] (a);
                    \vertex [above right=of a] (b);
                    \vertex [below right=of a] (c);
                    \vertex [above right=0.3cm and 0.6cm of b] (b1) {\!};
                    \vertex [below right=0.3cm and 0.6cm of b] (b2) {\!};
                    \vertex [above right=0.3cm and 0.6cm of c] (c1) {\!};
                    \vertex [below right=0.3cm and 0.6cm of c] (c2) {\!};
                    \diagram* {
                    (i1) -- [scalar] (a) -- [scalar,  edge label=$h_b$] (b),
                    (a) -- [scalar,edge label'=$h_c$] (c),
                    (b2) -- [fermion] (b) -- [fermion] (b1),
                    (c2) -- [fermion] (c) -- [fermion] (c1)
                    };
                    \draw [decoration={brace}, decorate] (b1.east) -- (b2.east)
                    node [pos=0.5, right] {$F_\text{SM}$};
                    \draw [decoration={brace}, decorate] (c1.east) -- (c2.east)
                    node [pos=0.5, right] {\(F_\text{SM}\)};
                \end{feynman}
            \end{tikzpicture}}\\
        \subfloat[single cascade\label{sfig:single}]{
            \begin{tikzpicture}
                \begin{feynman}
                    \vertex (i1) {\(h_{3} \)};
                    \vertex [right=of i1] (a);
                    \vertex [above right=of a] (b);
                    \vertex [below right=of a] (c);
                    \vertex [above right= 0.3cm and 0.6cm of b] (b1)  {\!};
                    \vertex [below right=0.3cm and 0.6cm of b] (b2)  {\!};
                    \vertex [above right=of c] (d);
                    \vertex [below right=of c] (e);
                    \vertex [above right=0.3cm and 0.6cm of d] (d1) {\!};
                    \vertex [below right=0.3cm and 0.6cm of d] (d2) {\!};
                    \vertex [above right=0.3cm and 0.6cm of e] (e1) {\!};
                    \vertex [below right=0.3cm and 0.6cm of e] (e2) {\!};
                    \diagram* {
                    (i1) -- [scalar] (a) -- [scalar, edge label=\(h_{1,2}\)] (b),
                    (a) -- [scalar, edge label'={\(h_2\)}] (c) -- [scalar, edge label={\(h_1\)},near end] (d),
                    (b2) -- [fermion] (b) -- [fermion] (b1),
                    (d2) -- [fermion] (d) -- [fermion, thick] (d1),
                    (c) -- [scalar, edge label'=$h_1$, near end] (e),
                    (e2) -- [fermion] (e) -- [fermion] (e1),
                    };
                    \draw [decoration={brace}, decorate] (b1.east) -- (b2.east)
                    node [pos=0.5, right] {\(F_\text{SM}\)};
                    \draw [decoration={brace}, decorate] (d1.east) -- (d2.east)
                    node [pos=0.5, right] {\(F_\text{SM}\)};
                    \draw [decoration={brace}, decorate] (e1.east) -- (e2.east)
                    node [pos=0.5, right] {\(F_\text{SM}\)};
                \end{feynman}
            \end{tikzpicture}}
    \end{minipage}
    \subfloat[double cascade\label{sfig:double}]{
        \begin{tikzpicture}
            \begin{feynman}
                \vertex (i1) {\(h_{3} \)};
                \vertex [right=of i1] (a);
                \vertex [above right = 1.7cm and 1cm of a] (b);
                \vertex [below right= 1.7cm and 1cm of a] (c);
                \vertex [above right= of b] (f);
                \vertex [below right=of b] (g);
                \vertex [above right=0.3cm and 0.6cm of f] (f1)  {\!};
                \vertex [below right=0.3cm and 0.6cm of f] (f2)  {\!};
                \vertex [above right=0.3cm and 0.6cm of g] (g1)  {\!};
                \vertex [below right=0.3cm and 0.6cm of g] (g2)  {\!};
                \vertex [above right=of c] (d);
                \vertex [below right=of c] (e);
                \vertex [above right=0.3cm and 0.6cm of d] (d1) {\!};
                \vertex [below right=0.3cm and 0.6cm of d] (d2) {\!};
                \vertex [above right=0.3cm and 0.6cm of e] (e1) {\!};
                \vertex [below right=0.3cm and 0.6cm of e] (e2) {\!};
                \diagram* {
                (i1) -- [scalar] (a)-- [scalar,  edge label=\(h_{2}\)] (b),
                (a) -- [scalar, edge label'={\(h_2\)}] (c) -- [scalar, edge label={\(h_1\)},near end] (d),
                (f) -- [scalar, edge label'={\(h_1\)}, near start] (b) -- [scalar, edge label'={\(h_1\)},near end] (g),
                (c) -- [scalar, edge label'=\(h_1\), near end] (e)
                (f2) -- [fermion] (f) -- [fermion] (f1),
                (g2) -- [fermion] (g) -- [fermion] (g1),
                (d2) -- [fermion] (d) -- [fermion] (d1),
                (e2) -- [fermion] (e) -- [fermion] (e1),
                };
                \draw [decoration={brace}, decorate] (f1.east) -- (f2.east)
                node [pos=0.5, right] {\(F_\text{SM}\)};
                \draw [decoration={brace}, decorate] (g1.east) -- (g2.east)
                node [pos=0.5, right] {\(F_\text{SM}\)};
                \draw [decoration={brace}, decorate] (d1.east) -- (d2.east)
                node [pos=0.5, right] {\(F_\text{SM}\)};
                \draw [decoration={brace}, decorate] (e1.east) -- (e2.east)
                node [pos=0.5, right] {\(F_\text{SM}\)};
            \end{feynman}
        \end{tikzpicture}}
    \caption{Possible Higgs-to-Higgs decay signatures involving three neutral
        (mass ordered) scalars $h_a$ ($a\in \{ 1,2,3\}$): (a) $h_a \to h_b h_c$
        (with $a>b,c$) with successive decay of $h_b$ and $h_c$ to SM particles;
        (b) $h_3 \to h_2 h_c$ (with $c \in \{1,2\}$) with successive decay
        $h_2\to h_1h_1$ and $h_k$ as well as all $h_1$ decaying to SM particles;
        (c) $h_3 \to h_2 h_2 \to h_1 h_1 h_1 h_1$ and all $h_1$ decaying to SM
        particles.}\label{fig:cascades}
\end{figure}

We will now turn to the more exotic
signatures resulting from \cref{eq:process_sym,eq:process_asym} that are not yet
under active investigation. Following the processes in \cref{eq:process_sym,eq:process_asym}, the
two produced scalar states may further decay directly to SM particles.
Alternatively, an $h_2$ final state may decay into the two lightest scalars:
$h_2 \to h_1h_1$. This can lead to interesting decay cascades leading to three
or four scalar states that eventually decay to SM particles. The possible decay
patterns within our model are depicted in a generic form in \cref{fig:cascades}.
Here and in the following we denote final states from Higgs decays composed of
SM particles (\ie gauge bosons or fermions) generically $F_\text{SM}$, unless
otherwise specified. For the more complicated final states we will use
$F^n_\text{SM}$ to denote an $n$-particle SM final state, where we count the SM
particles before their decay (i.e., $W^\pm$, $Z$, and $t$ are counted as one particle).
As discussed above, the relative fractions of their decay rates solely depend on
the mass of the decaying Higgs state.

We find that all possible Higgs-to-Higgs decay signatures,
\cref{eq:process_sym,eq:process_asym}, can appear at sizable rates in the
allowed TRSM parameter space. In the next section we therefore design six
two-dimensional benchmark scenarios that highlight these signatures in detail,
and are tailored to initiate dedicated experimental studies and facilitate the
design of corresponding searches. As a final remark, we briefly want to comment
on the possible size of the total width of the resonantly-produced scalar state
$h_a$. \Cref{fig:h125pairs}~(\emph{right}) shows the ratio of the total width
over the mass, $\Gamma_a/M_a$, in dependence of $M_a$ and the sum of $h_a$
decays to scalar states, $\mathrm{BR}(h_a\to \text{NP})$. Parameter points with
larger values of $\Gamma_a/M_a$ overlay parameter points with smaller values. We
can clearly see that parameter points with larger $\Gamma_a/M_a$ tend to have
sizable decay rates to scalar states. However, overall, $\Gamma_a/M_a$ never
exceeds values greater than around $18\%$ in the considered mass range up to
$\SI{1}{\TeV}$. In the discussion of the benchmark scenarios below we will
comment on cases where $\Gamma_a/M_a\gtrsim\SI{1}{\%}$.

\section{{Benchmark Scenarios}}
\label{sec:benchmarks}

In this section we define six benchmark scenarios in order to motivate and
enable dedicated experimental studies of Higgs-to-Higgs decay signatures. Each
scenario focusses on one (or more) novel signatures and features a (close-to)
maximal signal yield that can be expected within the model while obeying the
constraints described in \cref{sec:TRSMconstraints}. The benchmark scenarios are
defined as two-dimensional planes where all model parameters except for the two
non-$h_{125}$ scalar masses are fixed. A brief overview of the benchmark
scenarios is given in \cref{tab:benchmarkoverview}. For each benchmark scenario,
\textbf{BP1}--\textbf{BP6}, it specifies the Higgs state $h_a$ that is
identified with the observed Higgs state, $h_{125}$, the target Higgs-to-Higgs
decay signature, as well as the possibilities of phenomenologically
relevant\footnote{For instance, in BP2, the successive decay $h_2 \to h_1 h_1$
could in principle occur for the case that $M_1 < \SI{62.5}{\GeV}$, however,
Higgs signal rate measurements strongly constrain the possible decay rate, and
we do not further consider this possiblity here.} successive Higgs decays,
potentially leading to single or double cascade decay signatures (see
\cref{fig:cascades}).

\begin{table}
    \centering
    \begin{tabularx}{\textwidth}{sssb}
        \toprule
        benchmark scenario & $h_{125}$ candidate & target signature      & possible successive decays                          \\
        \midrule
        \textbf{BP1}       & $h_3$               & $h_{125} \to h_1 h_2$ & $h_2 \to h_1 h_1$ if $M_2 > 2 M_1$                  \\
        \textbf{BP2}       & $h_2$               & $h_3 \to h_1 h_{125}$ & -                                                   \\
        \textbf{BP3}       & $h_1$               & $h_3 \to h_{125} h_2$ & $h_2 \to h_{125}h_{125}$ if $M_2 > \SI{250}{\GeV}$  \\
        \textbf{BP4}       & $h_3$               & $h_2 \to h_1 h_1$     & -                                                   \\
        \textbf{BP5}       & $h_2$               & $h_3 \to h_1 h_1$     & -                                                   \\
        \textbf{BP6}       & $h_1$               & $h_3 \to h_2 h_2$     & $h_2 \to h_{125}h_{125}$ if  $M_2 > \SI{250}{\GeV}$ \\
        \bottomrule
    \end{tabularx}
    \caption{Overview of the benchmark scenarios: The second column denotes the
    Higgs mass eigenstate that we identify with the observed Higgs boson,
    $h_{125}$, the third column names the targeted decay mode of the resonantly
    produced Higgs state, and the fourth column lists possible relevant
    successive decays of the resulting Higgs states.}\label{tab:benchmarkoverview}
\end{table}

The model parameters for all scenarios as well as the coupling scale factors
$\kappa_a$ are given in \cref{tab:BPparams}. All cross section values given in
the following refer to production of the initial scalar through ggF at the
$\SI{13}{\TeV}$~LHC\@.

We employ a factorized approach relying on the narrow width approximation. For
each benchmark scenario we will show both the $\BR{h_a\to h_b h_c}$
($a,b,c \in \{ 1,2,3\}$, $a\neq b,c$) and the cross section
\begin{equation}
    \sigma(pp\to h_a \to h_b h_c) = \kappa^2_a \left.\sigma(gg\to h_\text{SM})\right|_{M_a}\cdot \BR{h_a\to h_b h_c}\eqdot
\end{equation}
In all scenarios where either $b=c$ or $h_{b,c}\equiv h_{125}$ there is only one
unknown BSM mass in the final state $h_b h_c$. In this case we will employ
a further factorization where we present the $\BR{h_b h_c \to F_\text{SM}}$ as a
function of the remaining mass parameter. In this case the full cross section
into a given SM final state can be obtained by
\begin{equation}
    \sigma(pp\to h_a \to h_b h_c \to F_\text{SM}) = \sigma(pp\to h_a \to h_b h_c) \cdot\BR{h_b h_c \to F_\text{SM}}\eqcomma
\end{equation}
where potential cascades, \cref{fig:cascades}, are included in the $\BR{h_b h_c
        \to F_\text{SM}}$ for $F^6_\text{SM}$ and $F^8_\text{SM}$.

All of the benchmark scenarios presented in the following are exemplary for the
corresponding signature within the TRSM\@. There are always alternative
choices for the fixed parameters that may lead to different cross sections,
branching ratios, or regions excluded by some constraints. As such, the regions
of parameter space that are excluded by some constraint in a benchmark scenario
should under no circumstances discourage experimental searches in this parameter
region.

\begin{table}[t]
    \centering
    \begin{tabularx}{\textwidth}{CRRRRRR}
        \toprule
        Parameter           & \multicolumn{6}{c }{Benchmark scenario}                                                                             \\
                            & \textbf{BP1}                            & \textbf{BP2} & \textbf{BP3} & \textbf{BP4} & \textbf{BP5} & \textbf{BP6}  \\
        \midrule
        $M_1~[\SI{}{\GeV}]$ & $[1, 62]$                               & $[1,124]$    & $125.09$     & $[1, 62]$    & $[1, 124]$   & $125.09$      \\
        $M_2~[\SI{}{\GeV}]$ & $[1, 124]$                              & $125.09$     & $[126, 500]$ & $[1,124]$    & $125.09$     & $[126, 500]$  \\
        $M_3~[\SI{}{\GeV}]$ & $125.09$                                & $[126,500]$  & $[255, 650]$ & $125.09$     & $[126, 500]$ & $[255, 1000]$ \\
        $\theta_{hs}$       & $1.435$                                 & $1.352$      & $-0.129$     & $-1.284$     & $-1.498$     & $0.207$       \\
        $\theta_{hx}$       & $-0.908$                                & $1.175$      & $0.226$      & $1.309$      & $0.251$      & $0.146$       \\
        $\theta_{sx}$       & $ -1.456$                               & $-0.407$     & $-0.899$     & $-1.519$     & $0.271$      & $0.782$       \\
        $v_s~[\SI{}{\GeV}]$ & $630$                                   & $120$        & $140$        & $990$        & $50$         & $220$         \\
        $v_x~[\SI{}{\GeV}]$ & $700$                                   & $890$        & $100$        & $310$        & $720$        & $150$         \\
        \midrule
        $\kappa_1$          & $0.083$                                 & $0.084$      & $0.966$      & $0.073$      & $0.070$      & $0.968$       \\
        $\kappa_2$          & $0.007$                                 & $0.976$      & $0.094$      & $0.223$      & $-0.966$     & $0.045$       \\
        $\kappa_3$          & $-0.997$                                & $-0.203$     & $0.239$      & $0.972$      & $-0.250$     & $0.246$       \\
        \bottomrule
    \end{tabularx}
    \caption{Input parameter values and coupling scale factors, $\kappa_a$
    ($a=1,2,3$), for the six defined benchmark scenarios. The doublet vev is set
    to $v=\SI{246}{\GeV}$ for all scenarios.}
    \label{tab:BPparams}
\end{table}

\subsection{BP1: \texorpdfstring{$h_{125}\to h_1 h_2$}{h125 -> h1 h2}}

\begin{figure}
    \centering
    \includegraphics[width=0.49\linewidth]{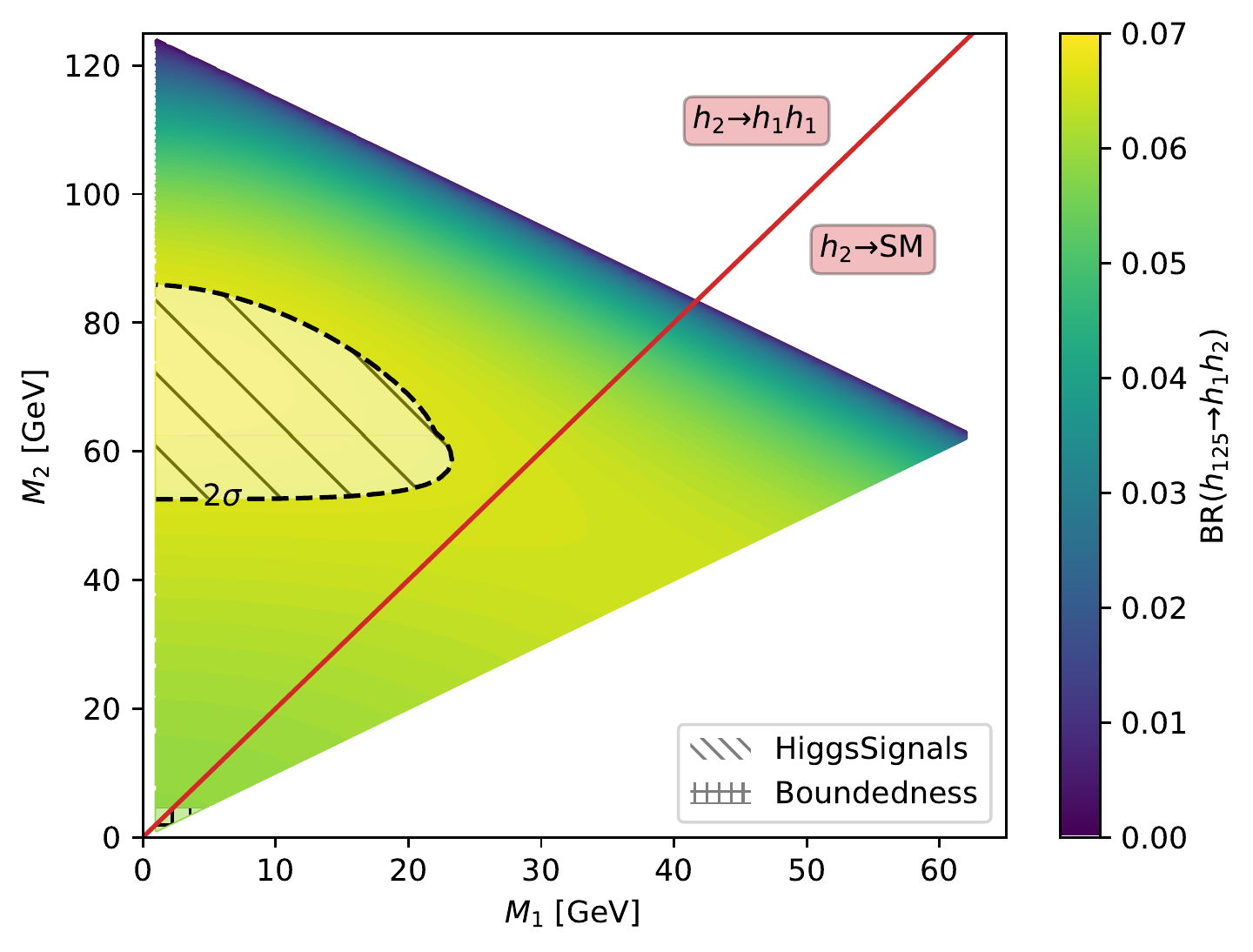}
    \includegraphics[width=.49\linewidth]{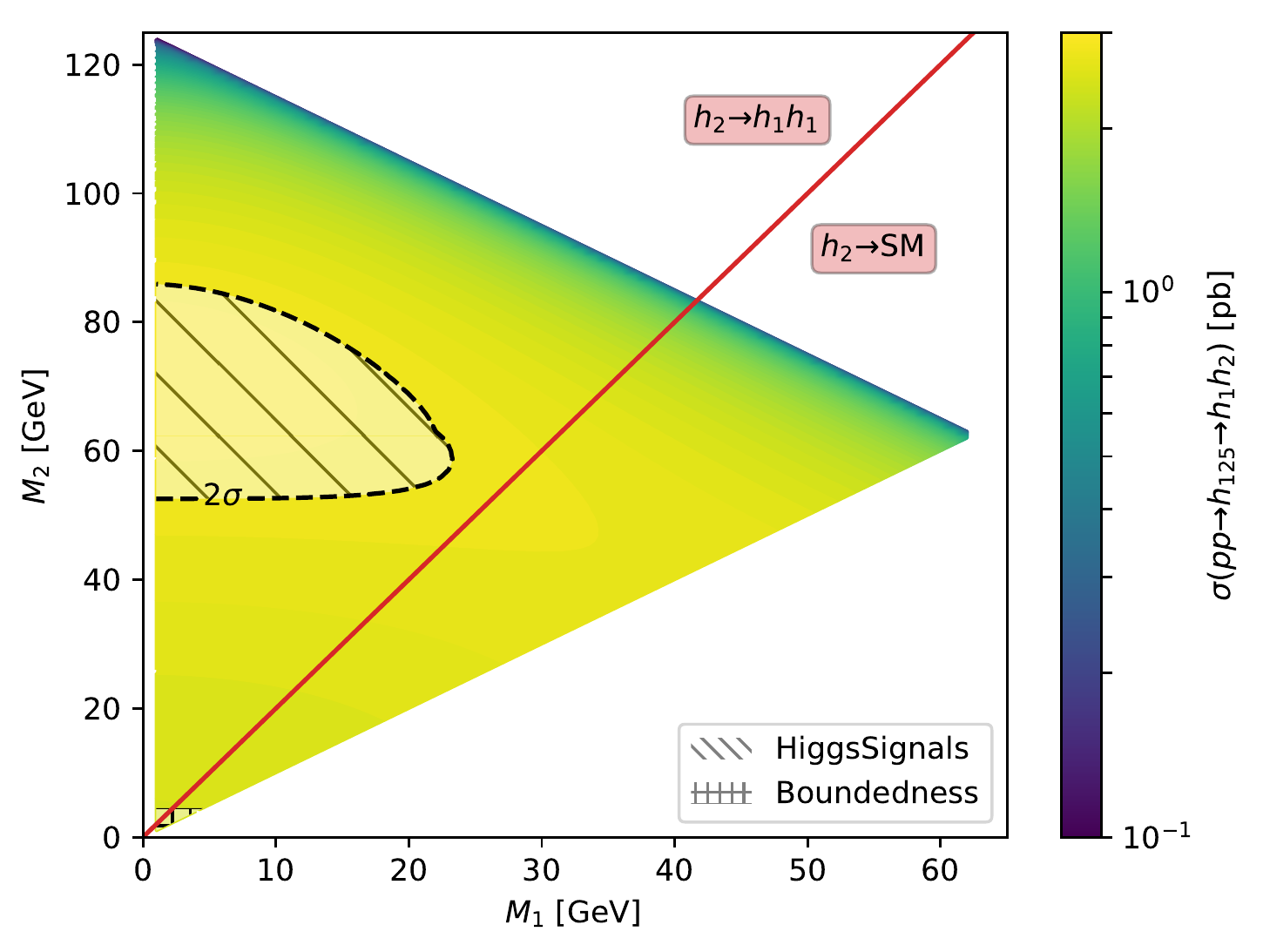}\\
    \includegraphics[width=.49\linewidth]{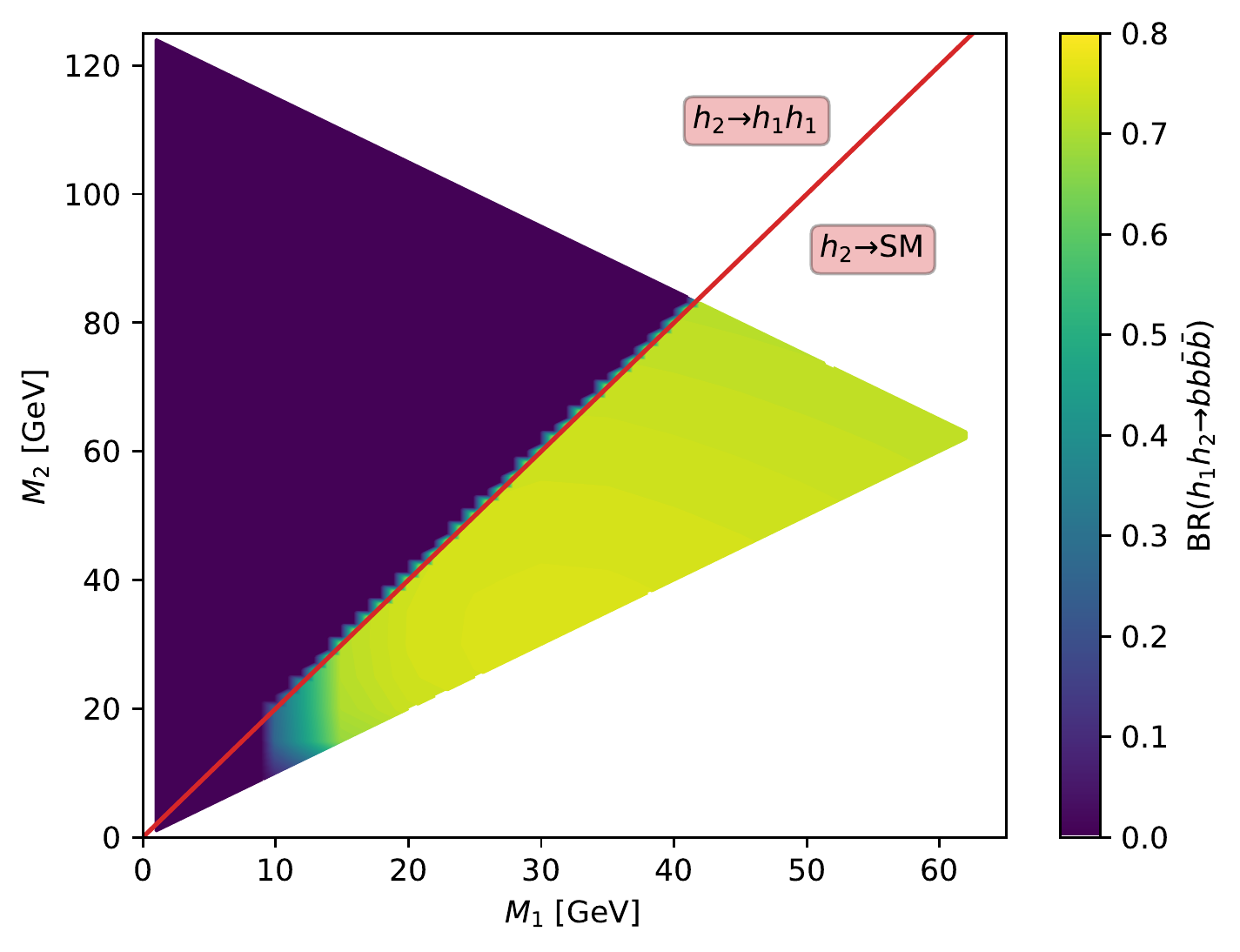}
    \includegraphics[width=.49\linewidth]{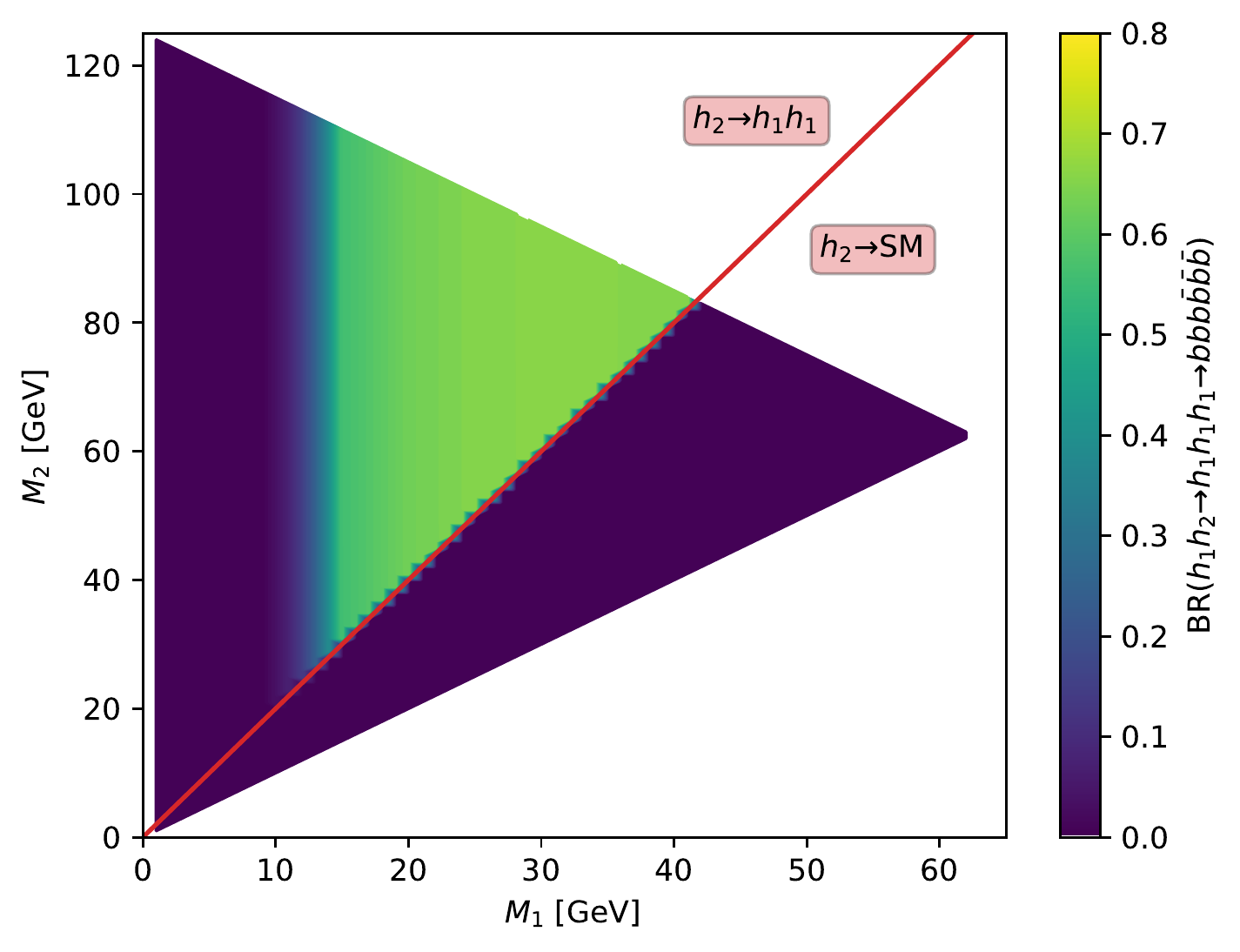}
    \caption{Benchmark plane \BP{1} for the decay signature $h_{125} \to h_1
            h_2$ with $h_{125}\equiv h_3$, defined in the $(M_1, M_2)$ plane.
        The color code shows $\BR{h_3 \to h_1 h_2}$ (\emph{top left panel}) and the \SI{13}{\TeV} LHC signal rate
        for $pp \to h_3 \to h_1 h_2$ (\emph{top right panel}). The red line separates the region
        $M_2 > 2 M_1$, where $\BR{h_2\to h_1 h_1}\approx 100\%$, from the
        region $M_2 < 2 M_1$, where $\BR{h_2 \to F_\text{SM}} \approx
            100\%$. The BR of the $h_1h_2$ state into
        $b\bar{b}b\bar{b}$ and --- through a $h_2\to h_1h_1$ cascade ---
        $b\bar{b}b\bar{b}b\bar{b}$ final states are shown in the \emph{bottom left} and \emph{right panels}, respectively.}\label{fig:bp1}
\end{figure}

In the first benchmark scenario, \BP{1}, we identify the heaviest scalar state
$h_3$ with $h_{125}$, and focus on the asymmetric decay $h_{125} \to h_1 h_2$.
The parameter values (see \cref{tab:BPparams}) are chosen such that the
couplings of $h_3$ to SM particles are nearly identical to the SM predictions,
$\kappa_3 \simeq 1$. At the same time, the parameter choice maximizes --- within
the experimentally allowed range ---  the branching ratio $\BR{h_{125} \to h_1
h_2}$, which is shown in \cref{fig:bp1} (\emph{top left}) as a function of $M_1$
and $M_2$. In \cref{fig:bp1} (\emph{top right}) we show the corresponding signal
rate for inclusive production via gluon gluon fusion. We find that the BR for
$h_3\to h_1h_2$ reaches up to $7-\SI{8}{\%}$ which translates into a signal rate
of around $\SI{3}{\pb}$. These maximal branching ratios are reached in the
intermediate mass range for $h_2$, $M_2\sim 60-\SI{80}{\GeV}$. This feature is
caused by the fact that the triple Higgs couplings are proportional to the
masses (see \cref{eq:asymmetriccoup}).  Therefore, although phase space opens up
significantly for light decay products, the branching ratios become smaller for
$ M_2 < \SI{40}{\GeV}$. In the hatched region in \cref{fig:bp1} the decay rate
slightly exceeds the $2\sigma$ upper limit inferred from the LHC Higgs rate
measurements (using \HiggsSignals). We stress again that this excluded area is
dependent on our parameter choices and strongly encourage experimental searches
to cover the whole mass range.

Due to the sum rule, \cref{eq:TRSMsumrule}, the coupling scale factors
$\kappa_{1,2}$ have to be very close to zero in order to achieve
$\kappa_3\sim1$. This means that the couplings of $h_1$ and $h_2$ to SM
particles are strongly suppressed. As a result, if the decay channel $h_2\to h_1
h_1$ is kinematically open, $M_2 > 2 M_1$, it is the dominant decay mode leading
to a significant rate for the $h_1h_1h_1$ final state. In \BP{1} we find that
$\BR{h_2\to h_1 h_1}\simeq \SI{100}{\%}$ in this kinematic regime (\ie above the
red line in \cref{fig:bp1}) with a very sharp transition at the threshold. If in
addition $M_1 \gtrsim \SI{10}{\GeV}$ the $h_1$ decays dominantly into $b\bar{b}$
leading to a sizeable rate for the $b\bar{b}b\bar{b}b\bar{b}$ final state as
shown in \cref{fig:bp1} (\emph{bottom right}).

If the $h_2 \to h_1 h_1$ decay is kinematically closed, $M_2 < 2 M_1$, both
scalars $h_1$ and $h_2$ decay directly to SM particles, with BRs identical to a
SM-like Higgs boson with the corresponding mass (see \cref{fig:SMBRs}).
Therefore, for masses $M_1, M_2 \gtrsim \SI{10}{\GeV}$,  the $b\bar{b}b\bar{b}$
final state dominates, as shown in \cref{fig:bp1}~(\emph{bottom left}), while at
smaller masses, combinations with $\tau$-leptons and eventually final states
containing charm quarks {and} muons become relevant.

\subsection{BP2: \texorpdfstring{$h_3 \to h_1 h_{125}$}{h3 -> h1 h125}}

In the second benchmark scenario, \BP{2}, we identify $h_{125}\equiv h_2$ and
consider the production of $h_3$ followed by the asymmetric decay $h_3\to
    h_1h_{125}$. The scenario is defined in the ($M_1, M_3$) parameter plane, and
the remaining parameters are fixed to the values given in \cref{tab:BPparams}.
The mixing angles are chosen such that the production rate of $h_3$ is
maximized, while the $h_2$ properties remain consistent with the measured Higgs
signal rates. This results in a $h_3$ production rate of roughly $4\%$ of the
production cross section for a $h_\text{SM}$ at the same mass.

\begin{figure}
    \centering
    \includegraphics[width=.49\linewidth]{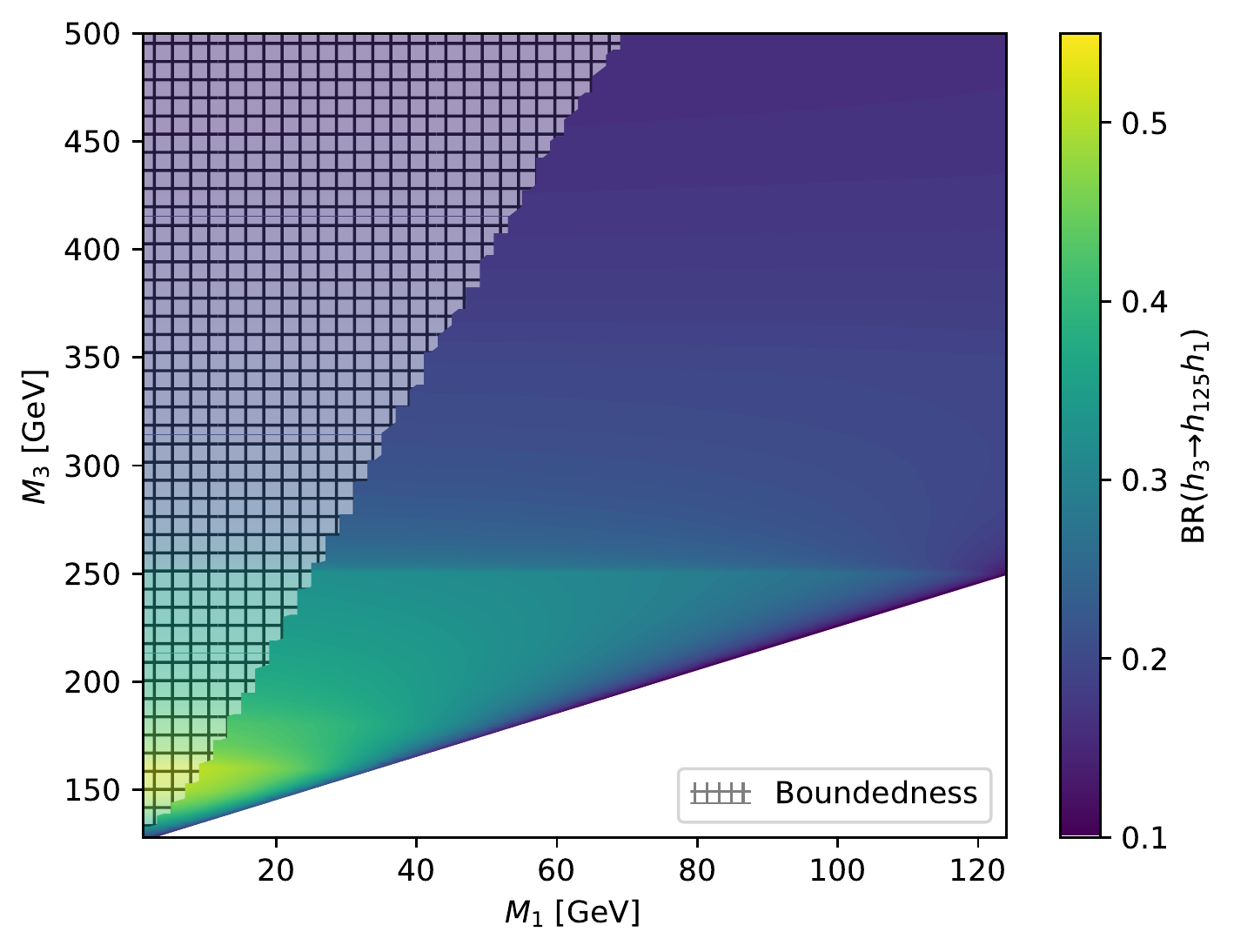}
    \includegraphics[width=.49\linewidth]{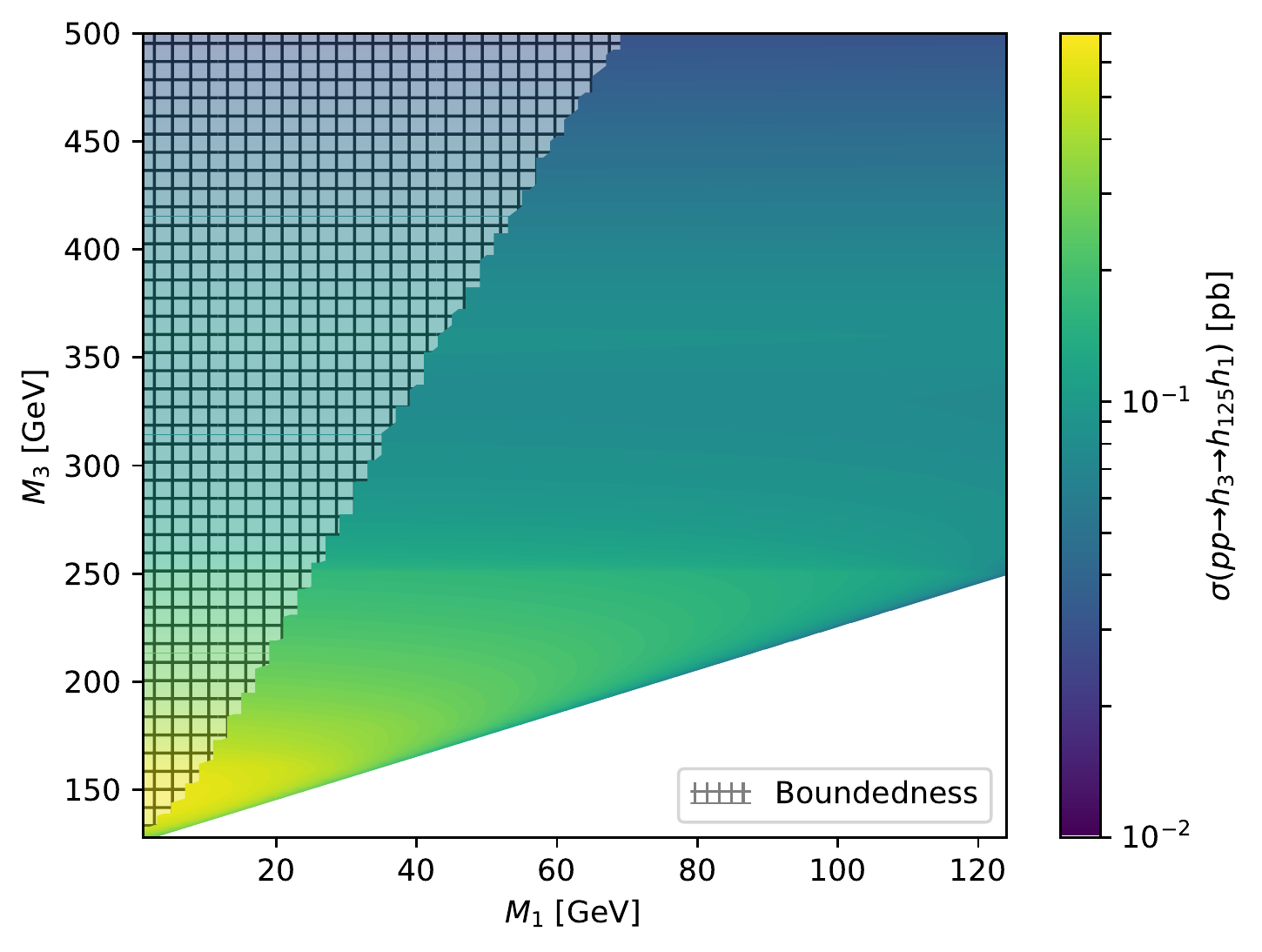}
    \caption{Benchmark plane \BP{2} for the decay signature $h_3 \to h_1
            h_{125}$ with $h_{125}\equiv h_2$, defined in the $(M_1, M_3)$
        plane. The color code shows $\BR{h_3 \to h_1 h_2}$ (\emph{left panel}) and the
        signal rate for $pp \to h_3 \to h_1 h_2$ (\emph{right panel}).}\label{fig:bp2}
\end{figure}

The phenomenology of \BP{2} is illustrated by \cref{fig:bp2}.  The $\BR{h_3\to
        h_1h_2}$ shown in \cref{fig:bp2} (\emph{left}) mostly stays above
$\SI{20}{\%}$ for $M_3 \lesssim \SI{350}{\GeV}$, reaching maximal values of
around $50-\SI{55}{\%}$ in the low mass region, $M_3 \sim 150 - \SI{170}{\GeV}$.
In this region, the corresponding signal rate in \cref{fig:bp2} (\emph{right})
is about $\SI{0.6}{\pb}$. It remains above $\SI{50}{\fb}$ as long as $M_3
    \lesssim \SI{450}{\GeV}$. The shaded region in \cref{fig:bp2} is excluded by
boundedness of the scalar potential. Again, this constraint depends strongly on
the values of the model parameters and should not discourage experimental
efforts to perform model-independent searches in this mass range. The total
    width of $h_3$ can reach maximal values of
    $\Gamma_3/M_3\sim\SI{1.1}{\%}$ in this benchmark scenario for
    $M_3\gtrsim \SI{480}{\GeV}$.

\begin{figure}
    \centering
    \includegraphics[width=0.53\linewidth]{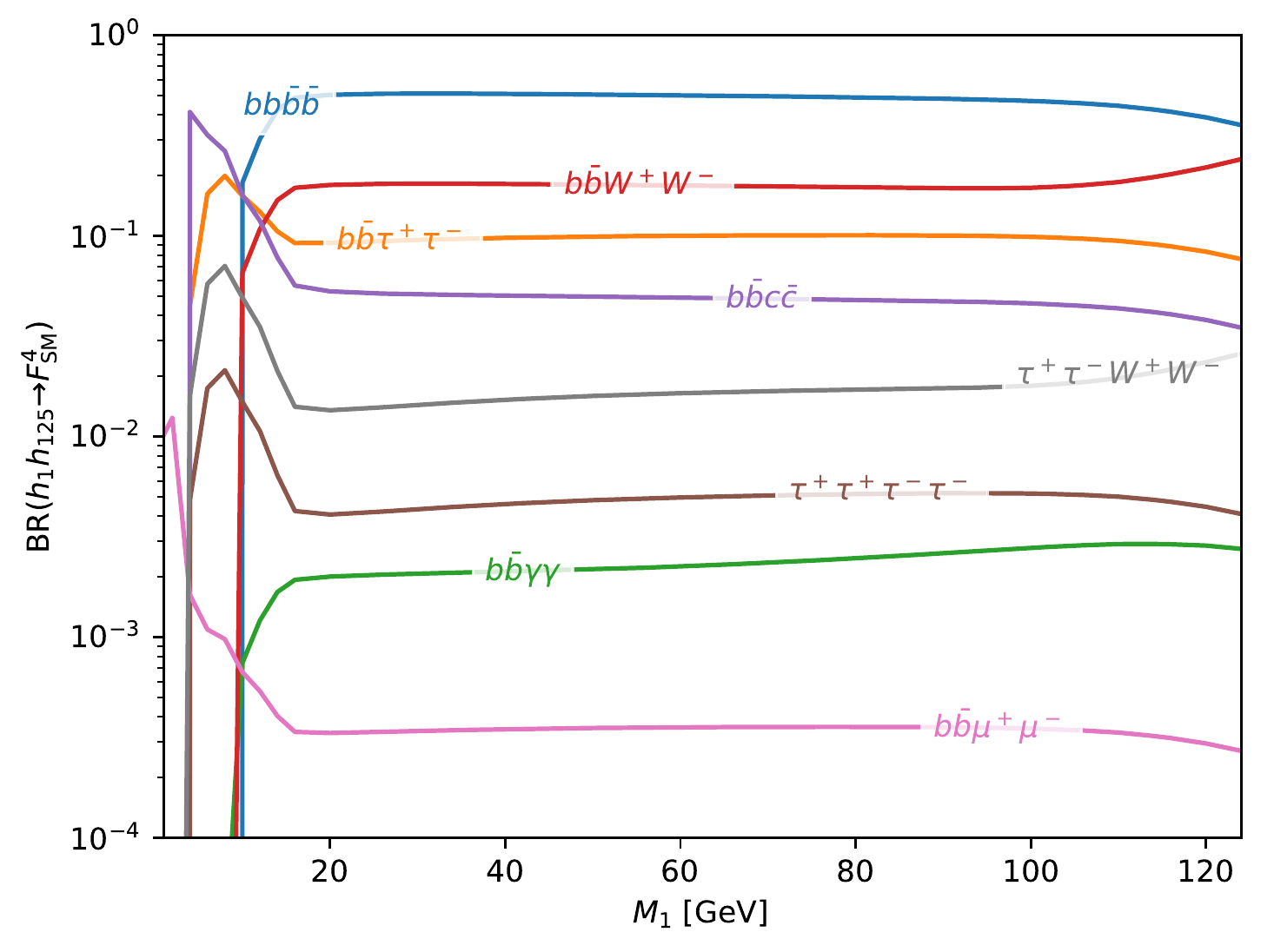}
    \caption{Branching ratios of the $h_1h_{125}$ state decaying into selected
        SM final states as a function of $M_1$ for \BP{2}.}\label{fig:BP2BRH1H2}
\end{figure}

The branching ratios for decays to SM final states originating from the $h_1h_{125}$
two-particle state are shown in \cref{fig:BP2BRH1H2} for \BP{2} as a function
of $M_1$. In most of the mass range, the
$b\bar{b}b\bar{b}$ final state dominates, followed by $b\bar{b} W^+W^-$ and
$b\bar{b} \tau^+\tau^-$ final states.

The cascade decay $h_{125}\equiv h_2\to h_1 h_1$ is in principle possible if
kinematically allowed and in compliance with the observed $h_{125}$ properties.
However, we chose $\kappa_2^2$ small in order to maximize $\kappa_3$ within the
experimental constraints. From \cref{fig:kappaBRNP} we see that, at the
corresponding value of $\kappa_2$, $\mathrm{BR}(h_{125} \to h_1h_1)$ must not
exceed $\sim\SI{2.5}{\%}$. In \BP{2} this decay rate is always below
$\SI{0.1}{\%}$.

Besides the asymmetric decay $h_3\to h_1 h_2$ the symmetric decays $h_3 \to h_1 h_1$ and $h_3\to h_2h_2$ are also present in this scenario. The decay $h_3 \to h_1h_1$ has a rate $\gtrsim \SI{25}{\%}$ in the mass range $M_3
    \lesssim\SI{250}{\GeV}$. The decay mode $h_3 \to h_2h_2$ only becomes
kinematically open for $M_3 \gtrsim 2 M_2 = \SI{250}{\GeV}$, and reaches
rates up to $\sim\SI{28}{\%}$. Although these rates are not negligible in
\BP{2}, we shall define dedicated benchmark scenarios \BP{5} and \BP{6} below where these decay modes clearly dominate.

\subsection{BP3: \texorpdfstring{$h_3\to h_{125}h_2$}{h3 -> h125 h2}}

\begin{figure}
    \centering
    \subfloat{\includegraphics[width=0.49\linewidth]{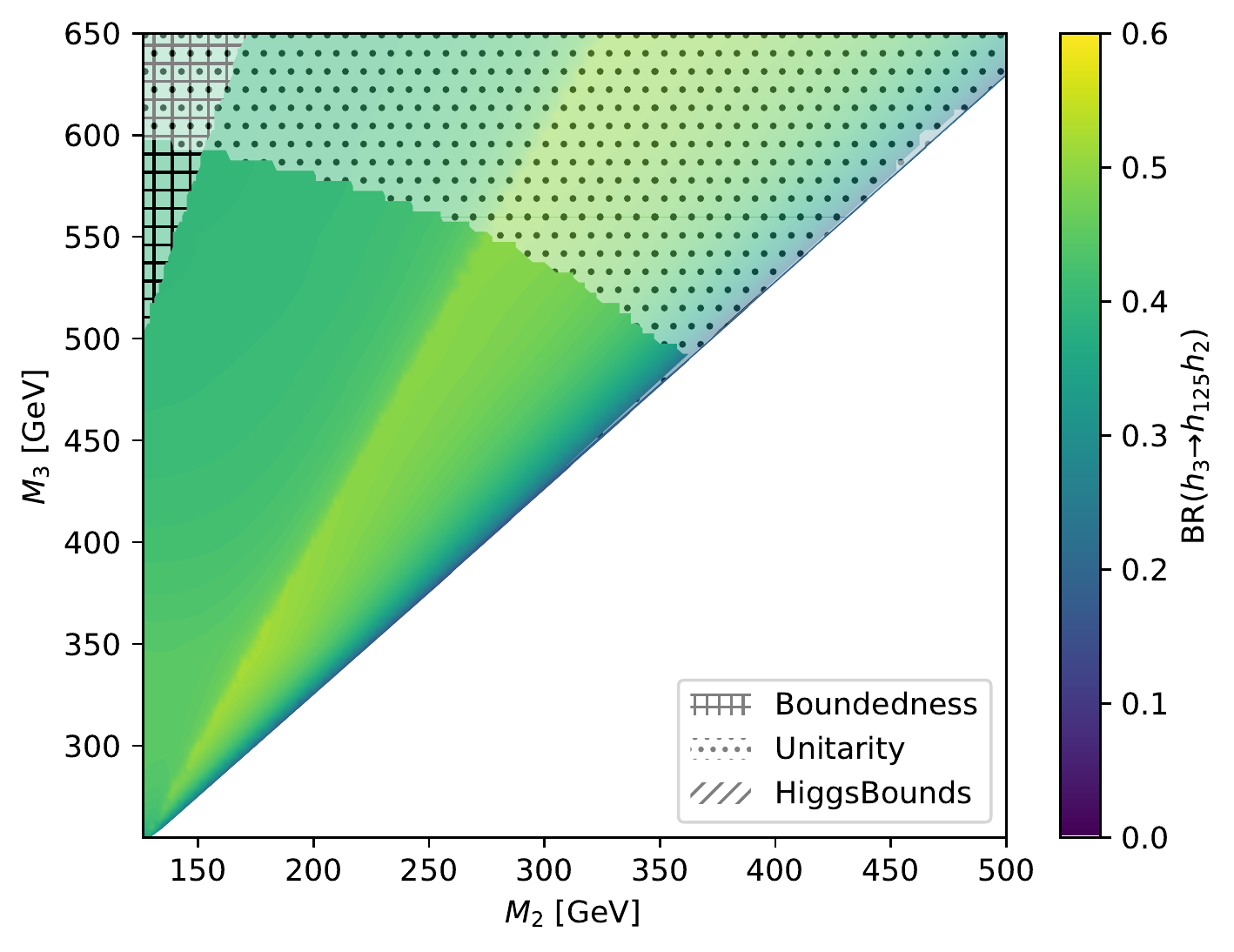}}
    \subfloat{\includegraphics[width=0.49\linewidth]{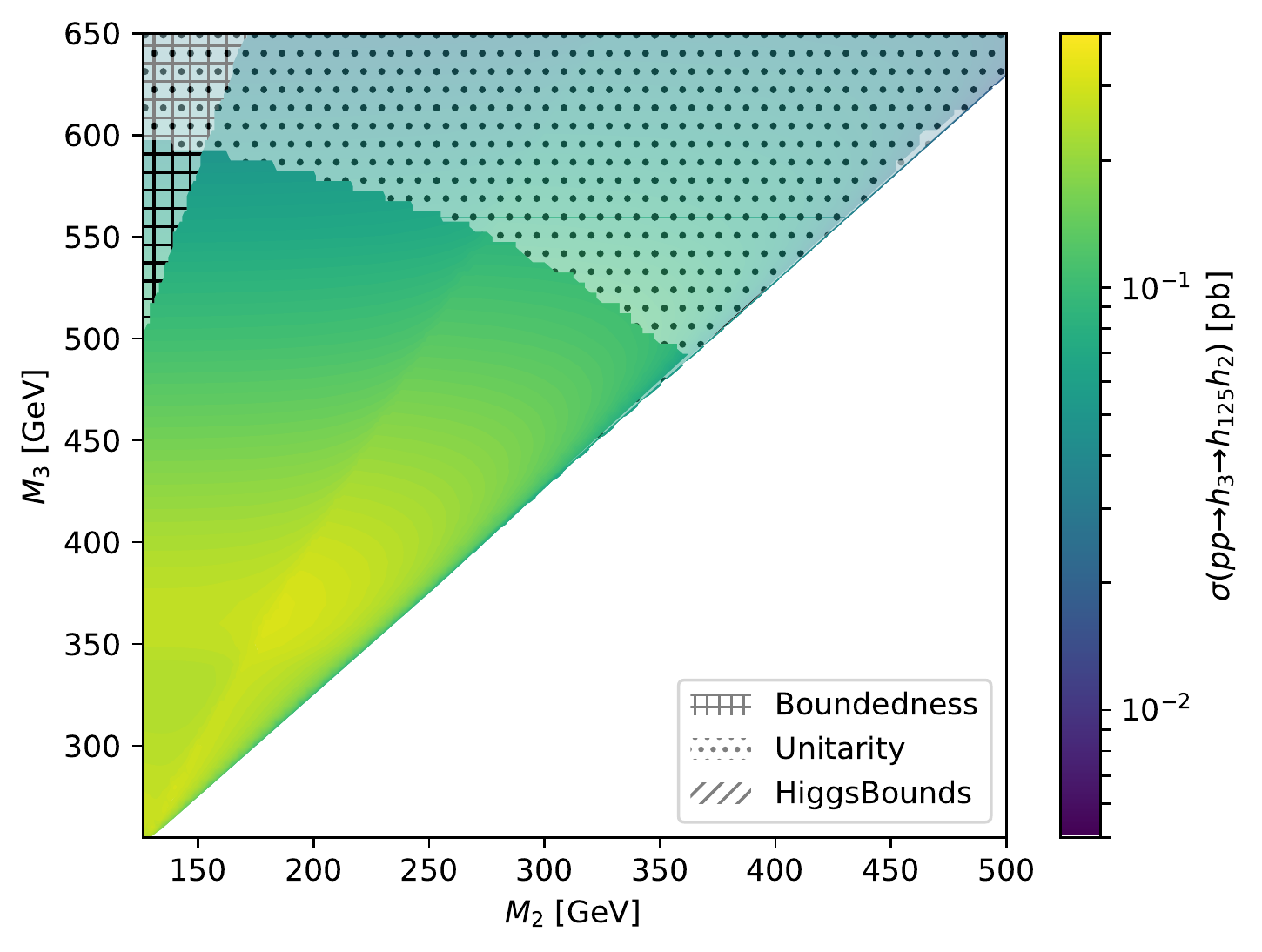}}
    \caption{Benchmark plane \BP{3} for the decay signature $h_3 \to h_{125}h_2$
        with $h_{125}\equiv h_1$, defined in the $(M_2, M_3)$ plane. The color
        code shows $\BR{h_3 \to h_{125} h_2}$ (\emph{left panel}) and the signal
        rate for $pp \to h_3 \to h_{125} h_2$ (\emph{right panel}). The shaded
        regions are excluded by boundedness from below, perturbative unitarity,
        and searches for heavy scalar resonances in diboson final
        states~\cite{Sirunyan:2018qlb,Aaboud:2018bun}.}\label{fig:bp3}
\end{figure}

In benchmark scenario \BP{3} we identify $h_{125}\equiv h_1$ and consider the
production of $h_3$ followed by the asymmetric decay $h_3\to h_{125}h_2$.
Similar to the \BP{2} scenario the mixing angles are chosen to maximize
$\kappa_3^2\simeq \SI{5.7}{\%}$ and $\BR{h_3\to h_1 h_2}$. The benchmark plane corresponding to the
parameters given in \cref{tab:BPparams} is shown in \cref{fig:bp3}.

The $\BR{h_3\to h_{125} h_2}$ shown in \cref{fig:bp3} (\emph{left}) is
$\gtrsim\SI{35}{\%}$ throughout the benchmark plane except for the region very
close to threshold. It reaches values around $\SI{50}{\%}$ in the parameter
region $M_3\lesssim 2 M_2$. The signal cross section, $\sigma(pp\to h_3 \to h_1
h_2)$ shown in \cref{fig:bp3} (\emph{right}), reaches up to $\SI{0.3}{\pb}$
while $M_3\lesssim\SI{500}{\GeV}$. At large values of $M_3 \gtrsim 500 -
\SI{600}{\GeV}$ the parameter space is partly constrained by perturbative
unitarity, and if simultaneously $M_1\lesssim\SI{150}{\GeV}$ the potential can
become unbounded from below, as indicated by the shaded regions. Very close to
its kinematic threshold, $M_3 \simeq  M_1 + \SI{125}{\GeV}$, the decay $h_3\to
h_{125}h_1$ is strongly suppressed. In this case, constraints can be derived
from current LHC searches for heavy resonances, in particular for the process
$pp \to h_3\to ZZ$~\cite{Sirunyan:2018qlb,Aaboud:2018bun}. The total width of
$h_3$ is maximal for the largest allowed values of $M_3$ and reaches
$\Gamma_3/M_3\sim\SI{4}{\%}$ for $M_3\gtrsim\SI{600}{\GeV}$.

\begin{figure}
    \centering
    \includegraphics[width=0.53\linewidth]{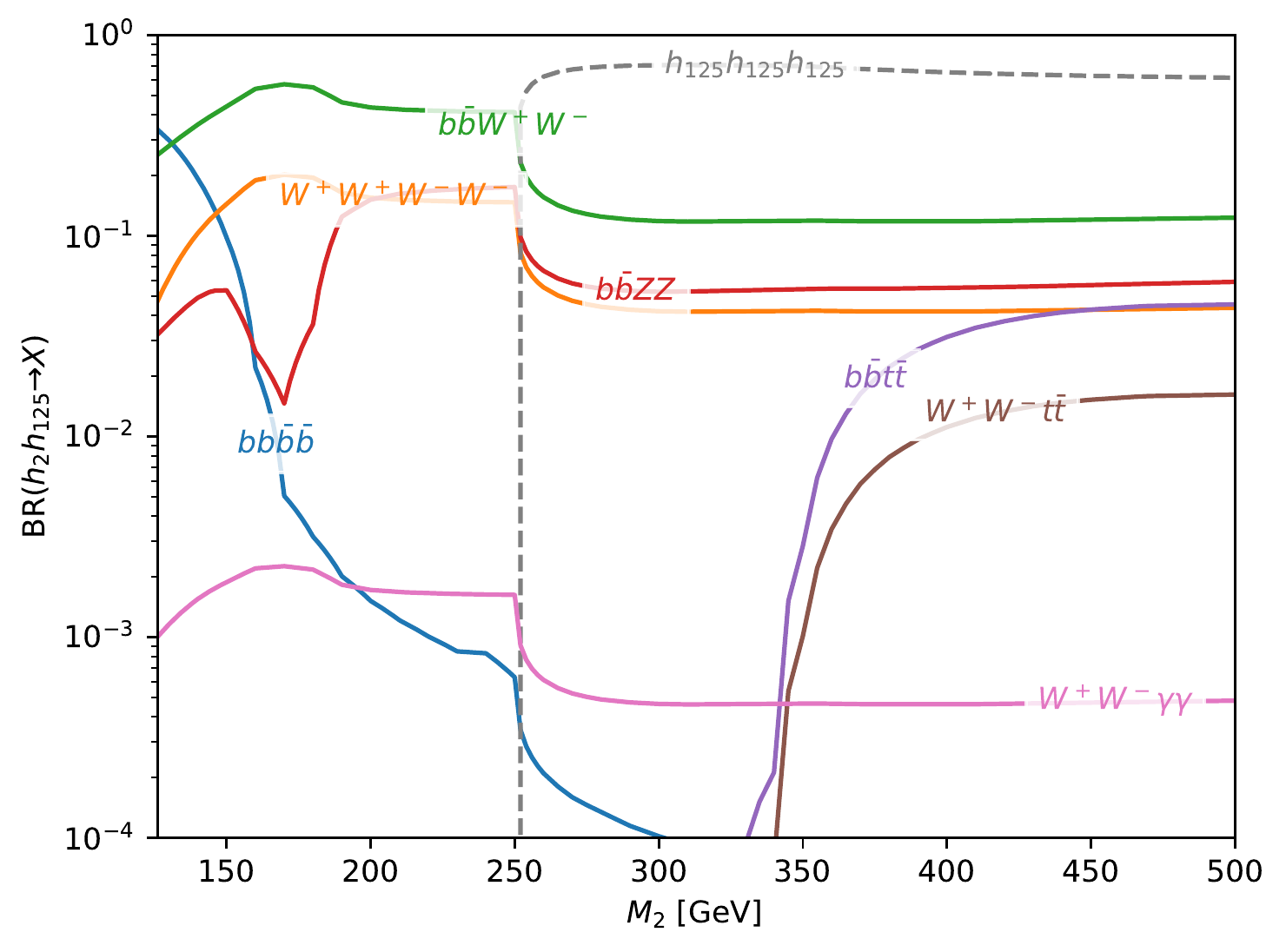}
    \caption{Branching ratios of the $h_{125}h_2$ state as a function of $M_2$
    for \BP{3}. Included are a selection of decay modes into SM particles as
    well as the cascade decay to $h_{125}h_{125}h_{125}$.}\label{fig:BP3BRH1H2}
\end{figure}

If $M_2<\SI{250}{\GeV}$ BSM decay modes of $h_2$ are prohibited and its decay
rates are identical to an $h_\text{SM}$ of the same mass (see \cref{fig:SMBRs}).
In this region the $h_{125}h_2$ state dominantly decays into final states
involving $b$-quarks and heavy gauge bosons as shown in \cref{fig:BP3BRH1H2}. As
soon as $M_2 > \SI{250}{\GeV}$ the decay $h_2\to h_{125}h_{125}$ becomes
dominant, quickly reaching a rate of $\sim\SI{70}{\%}$. Above threshold this
rate remains largely independent of $M_2$. The decay BRs of the resulting
$h_{125}h_{125}h_{125}$ state to the most important six particle SM final
states, $F^6_\text{SM}$, are given in \cref{tab:BP3triplebrs}. The first row
lists the direct branching ratios of $h_{125}h_{125}h_{125}$ while the second
row includes the factor $\BR{h_2\to h_{125}h_{125}}\approx\SI{68}{\%}$, which is
an approximation obtained in the mass region $\SI{260}{\GeV}<M_2 <
\SI{500}{\GeV}$. The resulting values can thus be compared directly to the BRs
of the four particle $F^4_\text{SM}$ in \cref{fig:BP3BRH1H2}. For instance, we
find that rates for $b\bar{b}b\bar{b}b\bar{b}$, $b\bar{b}b\bar{b}W^+ W^-$ and
$b\bar{b}W^+W^-$ final states are of comparable size for $M_2\gtrsim\SI{270}{\GeV}$.

\begin{table}
    \centering
    \begin{tabularx}{\textwidth}{bssssss}
        \toprule
        $\BR{X\to F^6_\text{SM}}$ & $6b$ & $4b\,2W$ & $2b\,4W$ & $4b\,2\tau$ & $4b\,2Z$ & $4b\,2\gamma$ \\
        \midrule
        $h_{125}h_{125}h_{125}$   & 20\% & 22\%     & 7.8\%    & 6.6\%       & 2.8\%    & 0.24\%        \\
        $h_2 h_{125}$             & 14\% & 15\%     & 5.3\%    & 4.5\%       & 1.9\%    & 0.16\%        \\
        \bottomrule
    \end{tabularx}
    \caption{{Decay rates of the $h_{125}h_{125}h_{125}$ state in \BP{3}, leading to a six-particle SM final state, $F_\text{SM}^6$. The second row gives the corresponding rates originating from the $h_2 h_{125}$ state, assuming $\BR{h_2\to{}h_{125}h_{125}}\approx\SI{68}{\%}$.}}
    \label{tab:BP3triplebrs}
\end{table}

The maximal production rates for the $h_3\to h_1 h_2 \to F_\text{SM}^4$ and
$h_3\to h_1 h_2 \to h_1h_1h_1 \to F_\text{SM}^6$ signatures amount to around
$\SI{0.3}{\pb}$ and $\SI{0.14}{\pb}$, respectively, where the latter is found
when both decays are just above threshold, $M_3 \simeq \SI{380}{\GeV}$ and $M_2
\simeq \SI{255}{\GeV}$.

In \BP{3}, the competing symmetric decay $h_3 \to h_2h_2$ reaches rates of
$\simeq \SI{20}{\%}$ if kinematically allowed. Otherwise the decay $h_3 \to
h_{125}h_{125}$ reaches similar values (and becomes dominant in the threshold
region, $M_3 \sim M_1 + M_2$).

\subsection{BP4: \texorpdfstring{$h_2\to h_1 h_1$ with $h_{125}\equiv h_3$}{h2 -> h1 h1 with h125=h3}}
We now turn to the symmetric Higgs-to-Higgs decay signatures. In benchmark
scenario \BP{4} we identify $h_{125}\equiv h_3$ and focus on the production of
$h_2$ followed by its decay $h_2\to h_1h_1$. In order to avoid constraints from
the Higgs rate measurements on the possible decays $h_{125} \to h_a h_b$ ($a,b
\in \{1,2\}$), the relevant couplings must be tuned to rather small values while
keeping $|\kappa_2|$ relatively large to ensure sizeable direct production of
$h_2$. The parameter choices for \BP{4} are listed in \cref{tab:BPparams}.

\begin{figure}
    \centering
    \subfloat{ \includegraphics[width=.49\linewidth]{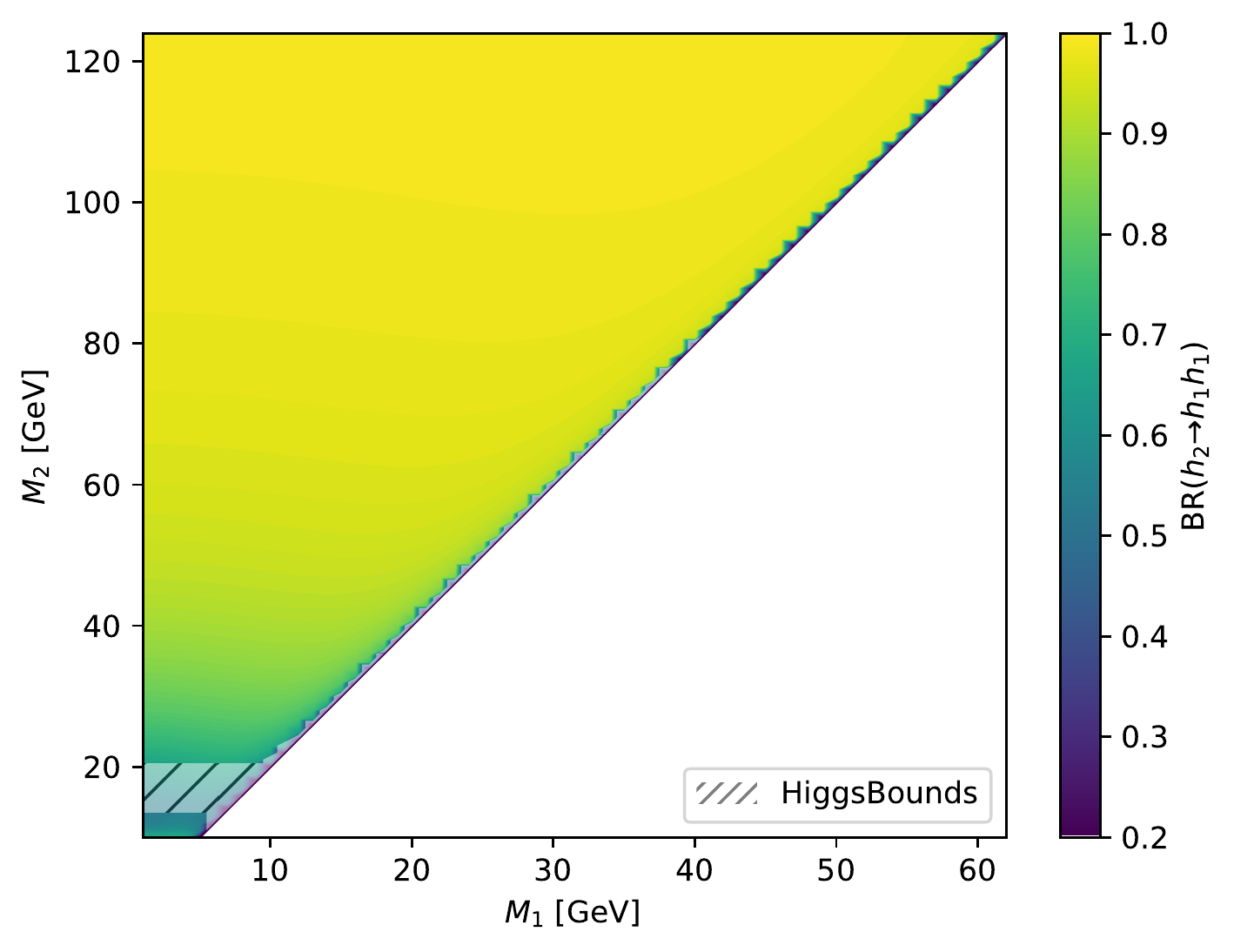}}
    \subfloat{ \includegraphics[width=.49\linewidth]{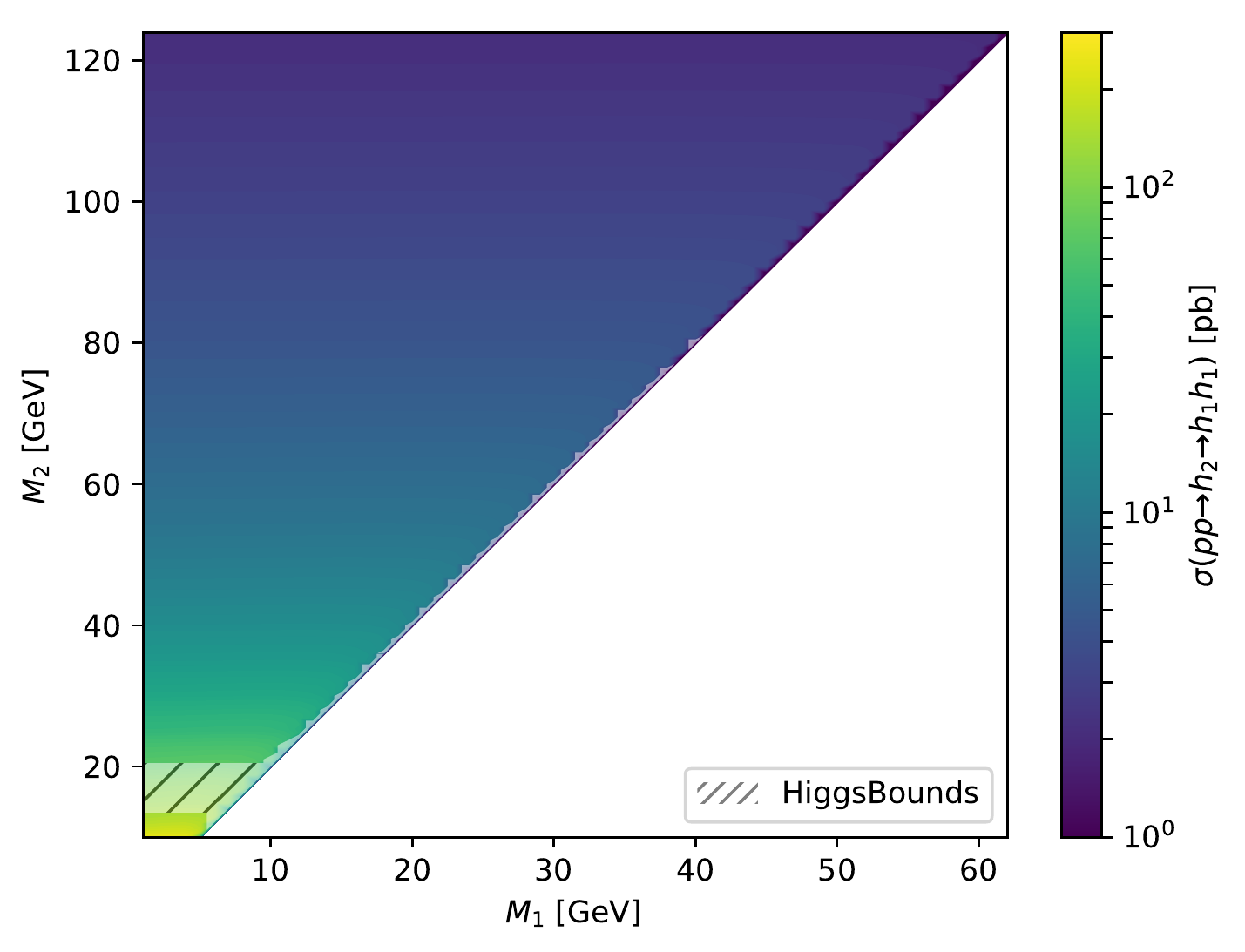}}
    \caption{Benchmark plane \BP{4} for the decay signature $h_2 \to h_1h_1$
        with $h_{125}\equiv h_3$, defined in the $(M_1, M_2)$ plane. The color code
        shows $\BR{h_2 \to h_1h_1}$ (\emph{left panel}) and the signal rate for $pp \to h_2 \to
            h_1h_1$ (\emph{right panel}). The shaded region is excluded by LEP searches for $e^+e^- \to Z
            h_2 \to Z (b \bar b)$~\cite{Schael:2006cr}.}\label{fig:bp4}
\end{figure}

\cref{fig:bp4} shows the collider phenomenology of $\text{BP4}$.
The branching ratio $\BR{h_2\to h_1h_1}$ is larger than $\SI{50}{\%}$ throughout
the allowed parameter plane, as shown in \cref{fig:bp4} (\emph{left}). For $M_2\gtrsim
    \SI{40}{\GeV}$ it is by far the dominant decay mode of $h_2$ with a BR of
more than $\SI{90}{\%}$. As the produced scalar boson is light, the signal rates shown in
\cref{fig:bp4} (\emph{right}) are enhanced by the large ggF cross section for light
scalars. Even though $h_2$ is only produced with a rate of about
$\kappa_2^2\sim5\%$ of the SM Higgs cross section at the same mass, we still
obtain signal rates of $\mathcal{O}(\SI{100}{\pb})$ in the low mass region $M_2
    \lesssim\SI{20}{\GeV}$. However, this parameter region is partly constrained by
LEP searches for $e^+e^- \to Z h_2 \to Z (b \bar{b})$~\cite{Schael:2006cr}. For
$M_2\ge\SI{20}{\GeV}$, where this limit is no longer sensitive, the signal rate
can still reach $\SI{60}{\pb}$. Still, the signature remains experimentally
challenging as the decay products for these low $M_1$ will be very soft.

\begin{figure}
    \centering
    \includegraphics[width=0.53\linewidth]{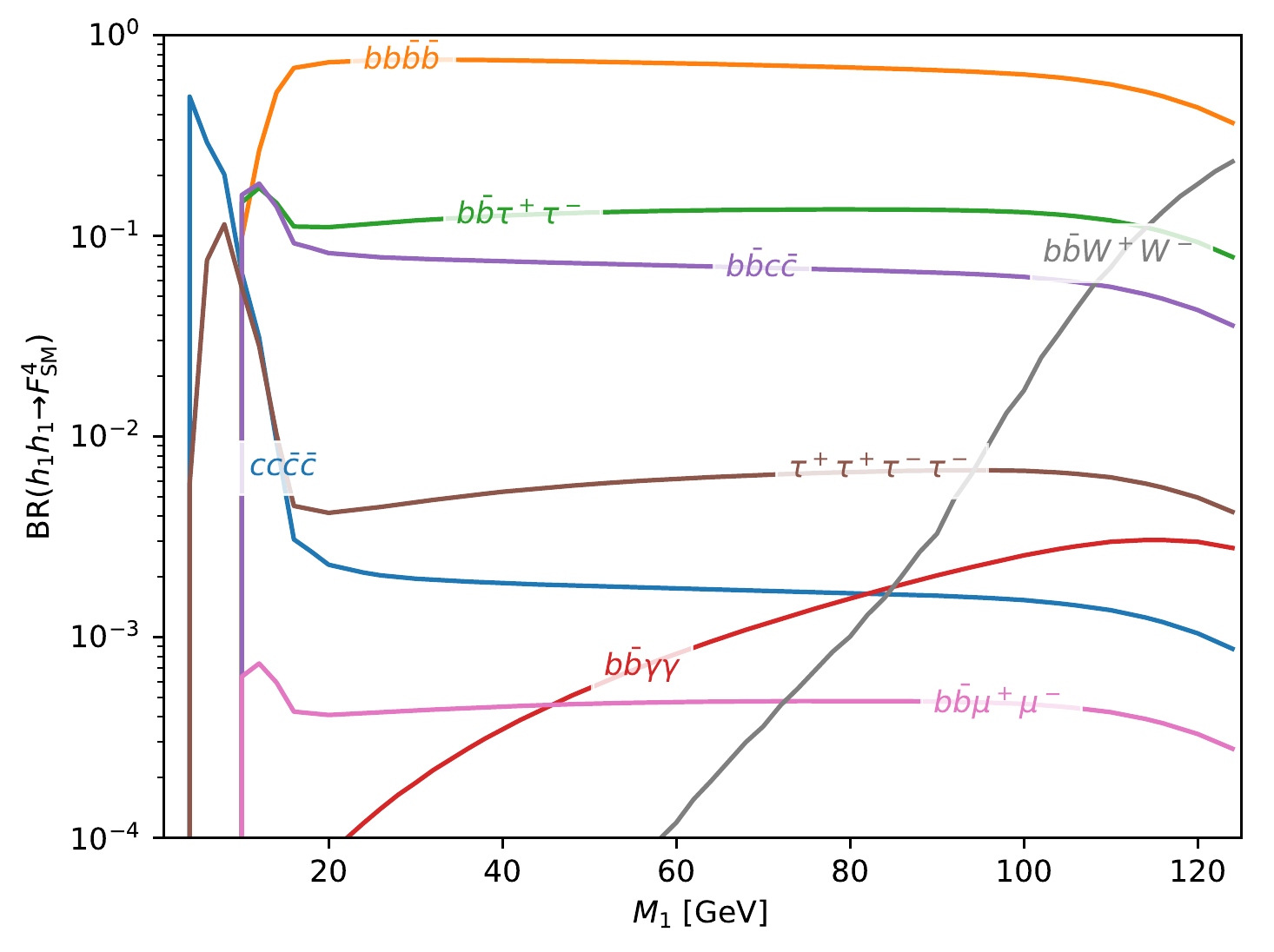}
    \caption{{Branching ratios} of the $h_1h_1$ state of \BP{4} and \BP{5} into selected SM decay modes
        as a function of $M_1$.}\label{fig:BRH1H1}
\end{figure}

The BRs for the decay modes of the $h_1h_1$ state into SM particles
are shown in \cref{fig:BRH1H1}. For $M_1 \gtrsim \SI{10}{\GeV}$ the decay into
$b\bar{b}b\bar{b}$ is dominant, followed by $b\bar{b}\tau^+\tau^-$. For even
lighter $M_1$ the predominant decay is into charm quarks.

The $h_{125}h_1h_1$ coupling is very small in this scenario. Still --- due to
the large $\kappa_{125}$ --- the process $p p \to h_{125}\to h_1 h_1$ can
enhance the total $h_1 h_1$ production cross section by up to $\sim 15\%$ for
large $M_2\sim\SI{125}{\GeV}$. On the other hand, interference effects between
the resonant $h_2$ and $h_{125}$ contributions --- similar to those discussed in
Ref.~\cite{Basler:2019nas} --- remain negligible.

\subsection{BP5: \texorpdfstring{$h_3\to h_1 h_1$ with $h_{125}\equiv h_2$}{h3 -> h1 h1 with h125=h2}}
I
n the benchmark plane \BP{5} we identify $h_{125}\equiv h_2$ and consider the
production of the heavier scalar $h_3$ followed by its symmetric decay to the
lightest scalar, $h_3\to h_1 h_1$. In our parameter scan of the TRSM (see
\cref{sec:TRSMscan}) we found that parameter points exhibiting a sizeable $pp\to
    h_3 \to h_1 h_1$ rate also tend to be strongly constrained by the Higgs signal
strength measurements if $2 M_1 < \SI{125}{\GeV}$. In addition, if kinematically
accessible, the decay modes $h_3\to h_2 h_2$ and/or $h_3 \to h_1 h_2$ tend to
dominate over the decay $h_3 \to h_1 h_1$. In order to define a suitable
benchmark scenario for the $pp\to h_3 \to h_1 h_1$ process it is therefore
necessary that all triple Higgs couplings except for $\tilde{\lambda}_{113}$ are
small while not overly suppressing $\kappa_3$. The chosen parameter values of
\BP{5} are given in \cref{tab:BPparams}.

\begin{figure}
    \centering
    \subfloat{\includegraphics[width=.49\linewidth]{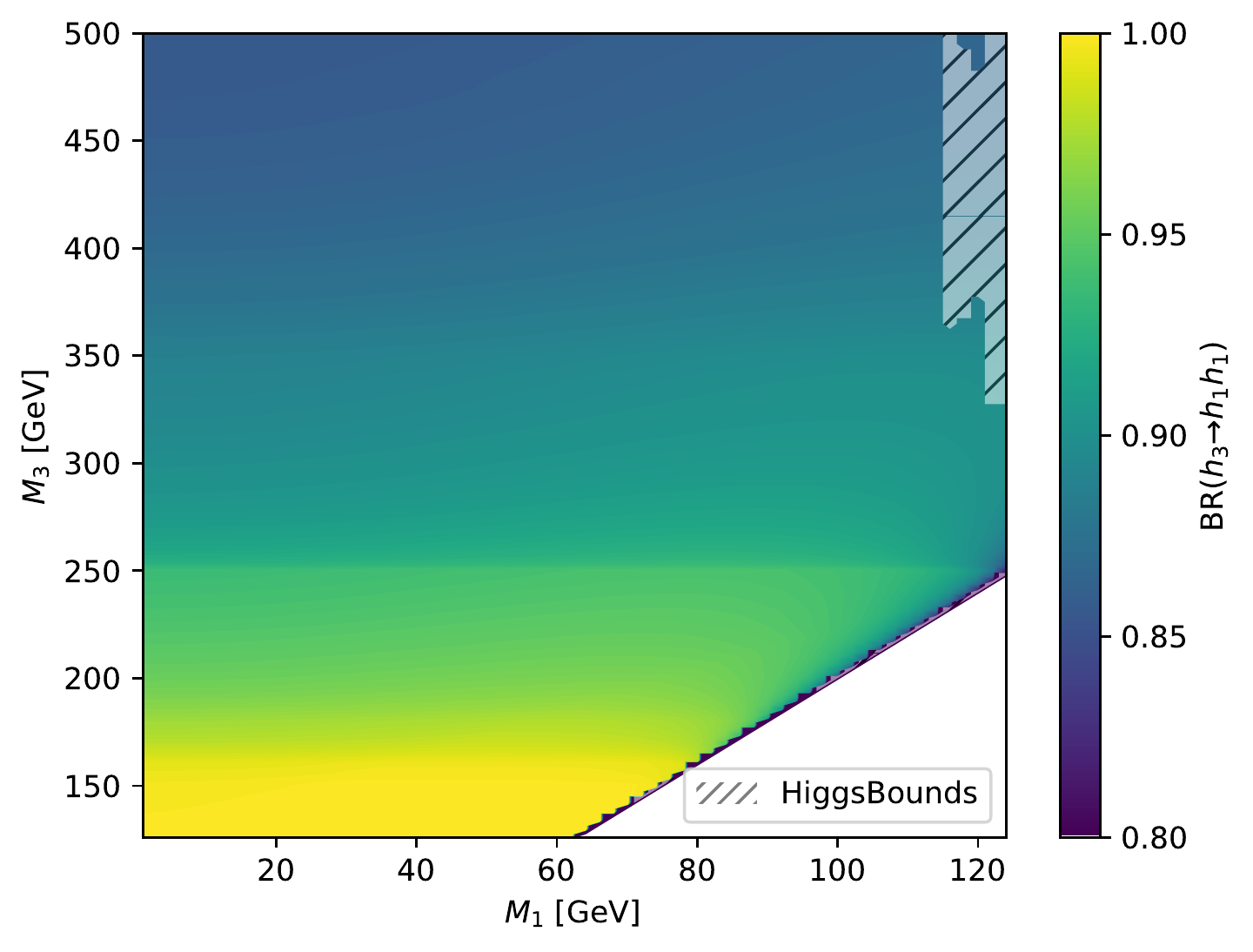}}
    \subfloat{\includegraphics[width=.49\linewidth]{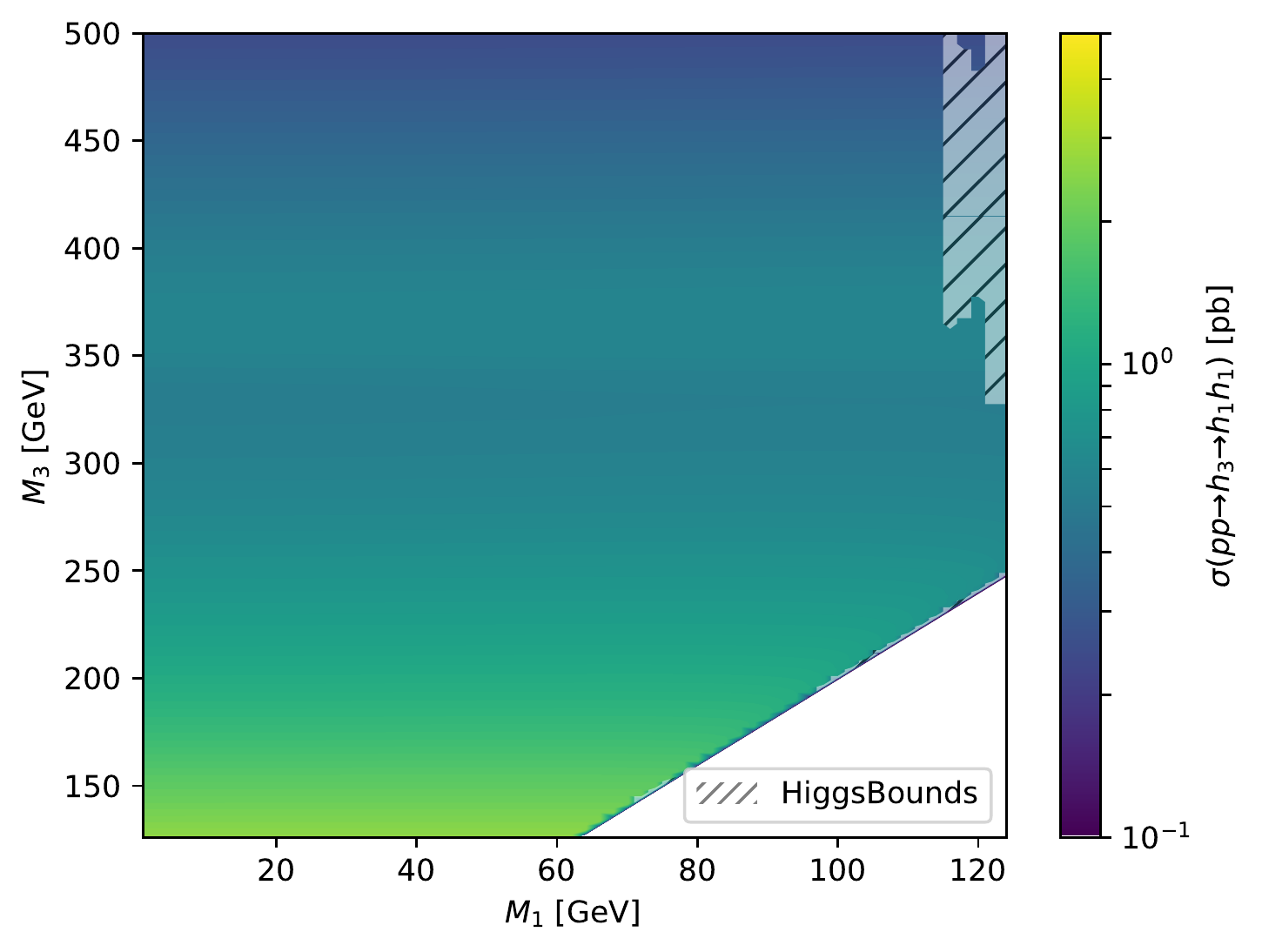}}
    \caption{Benchmark plane \BP{5} for the decay signature $h_3 \to h_1h_1$
        with $h_{125}\equiv h_2$, defined in the $(M_1, M_3)$ plane. The color code
        shows $\BR{h_3 \to h_1h_1}$ (\emph{left panel}) and the signal rate for $pp \to h_3 \to
            h_1h_1$ (\emph{right panel}). The shaded region is excluded by searches for resonant double
        Higgs production~\cite{Aaboud:2018sfw,Sirunyan:2018two} via
        \HiggsBounds.}\label{fig:bp5}
\end{figure}

The phenomenology of \BP{5} is shown in \cref{fig:bp5}. Throughout the parameter
plane $\BR{h_3\to h_1 h_1}$ --- shown in \cref{fig:bp5} (\emph{left}) ---
exceeds $\SI{85}{\%}$ and approaches $\SI{100}{\%}$ for low values of $M_3$. The
heavy scalar $h_3$ is produced at a rate of around $\kappa_3^2\simeq \SI{6}{\%}$
of the corresponding prediction for a SM Higgs boson. \Cref{fig:bp5}
(\emph{right}) shows the resulting signal rates of $\mathcal{O}(0.1 -
    \SI{1}{\pb})$ with maximal values of around $\SI{3}{\pb}$ for light $M_3\lesssim
    \SI{150}{\GeV}$. The parameter region at $M_1 \gtrsim \SI{120}{\GeV}$ and $M_3
    \gtrsim \SI{350}{\GeV}$ is constrained by LHC Higgs searches for resonant double
Higgs production~\cite{Aaboud:2018sfw,Sirunyan:2018two}. These are applied under
the assumption that $h_1$ cannot be experimentally distinguished from
$h_{125}\equiv h_2$ if they are close in mass and thus contributes to the
predicted signal rate for this process.

The BRs of the $h_1h_1$ two-particle state can again be found in
\cref{fig:BRH1H1}. They are identical to those discussed for \BP{4} since the
BRs of $h_1$ are always identical to those of a SM Higgs boson of the same mass
(see \cref{sec:TRSMcollpheno}). However, now the scenario extends to $M_1$
values up to $\SI{125}{\GeV}$ and with increasing $M_1$ the final state
$b\bar{b}W^+W^-$ becomes sizable. In contrast to \BP{4}, the two $h_1$ may be
boosted if $M_3 \gg 2 M_1$, leading to collimated $h_1$ decay products. This may
provide an additional experimental handle, enabling the reduction of
combinatoric background, and leading to a potential increase of the trigger
sensitivity as well as the applicability of jet substructure techniques.
Indeed, a recent ATLAS search for highly collimated photon-jets~\cite{Aaboud:2018djx} probes the signature $pp \to h_3 \to h_1h_1 \to \gamma\gamma\gamma\gamma$ in the mass range $M_3 \ge \SI{200}{\GeV}$, $ \SI{0.1}{\GeV} \le M_1 \le \SI{10}{\GeV}$. However, the currently obtained limit is still several orders of magnitude larger than the predicted rate in $\BP{5}$.

\subsection{BP6: \texorpdfstring{$h_3\to h_2 h_2$ with $h_{125}\equiv h_1$}{h3 -> h2 h2 with h125=h1}}

In benchmark plane \BP{6} we identify $h_{125}\equiv h_1$ and consider the
production of the heaviest scalar $h_3$ followed by its symmetric decay $h_3\to
    h_2h_2$. This constrains the mass range for $h_3$ to values $M_3 > \SI{250}{\GeV}$.
This, in combination with the suppression of $\kappa_3$ due to the sum rule,
\cref{eq:TRSMsumrule}, leads to relatively low production cross sections. The
input parameters for \BP{6} are listed in \cref{tab:BPparams}.

\begin{figure}
    \centering
    \includegraphics[width=.49\linewidth]{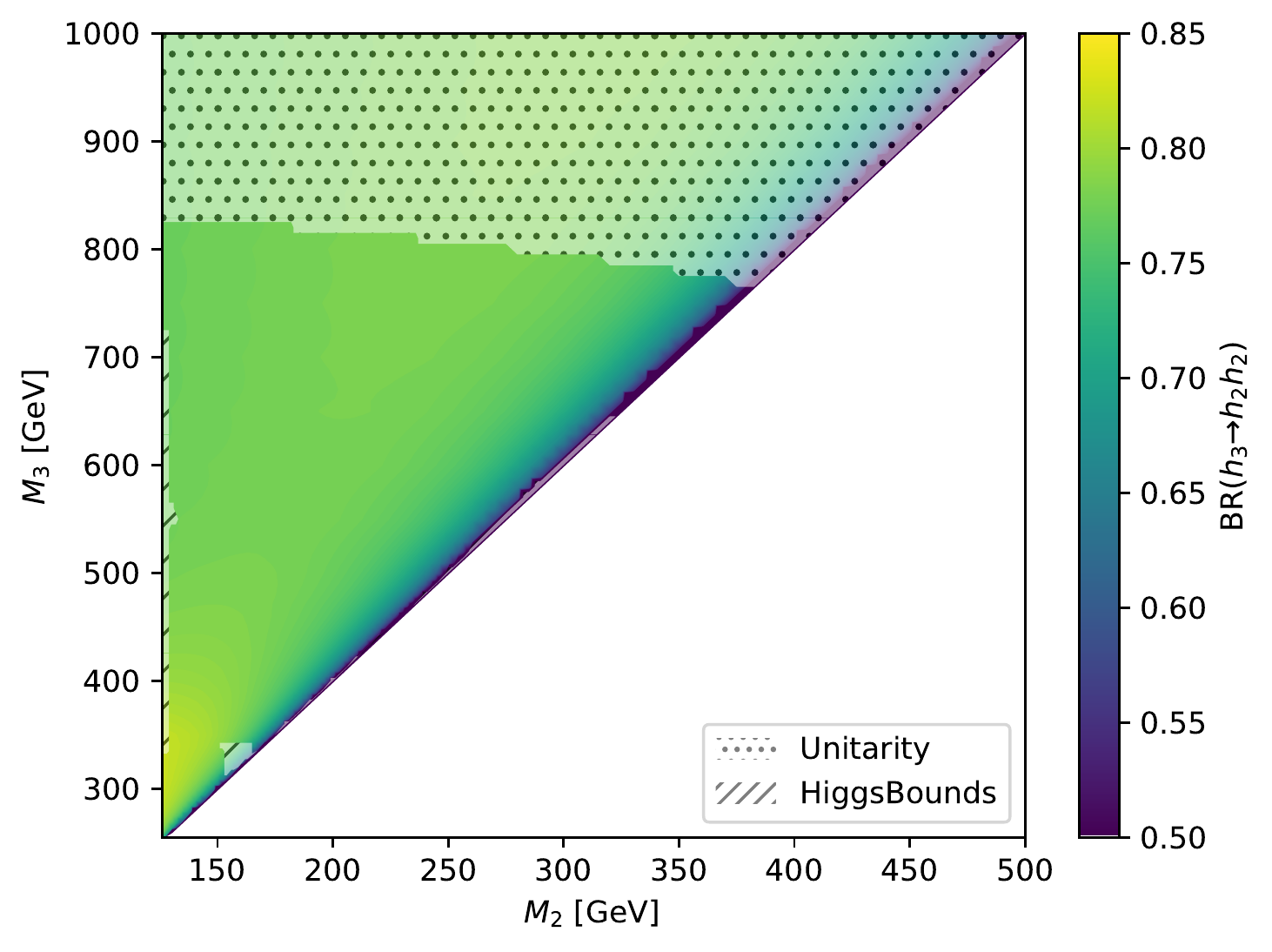}
    \includegraphics[width=.49\linewidth]{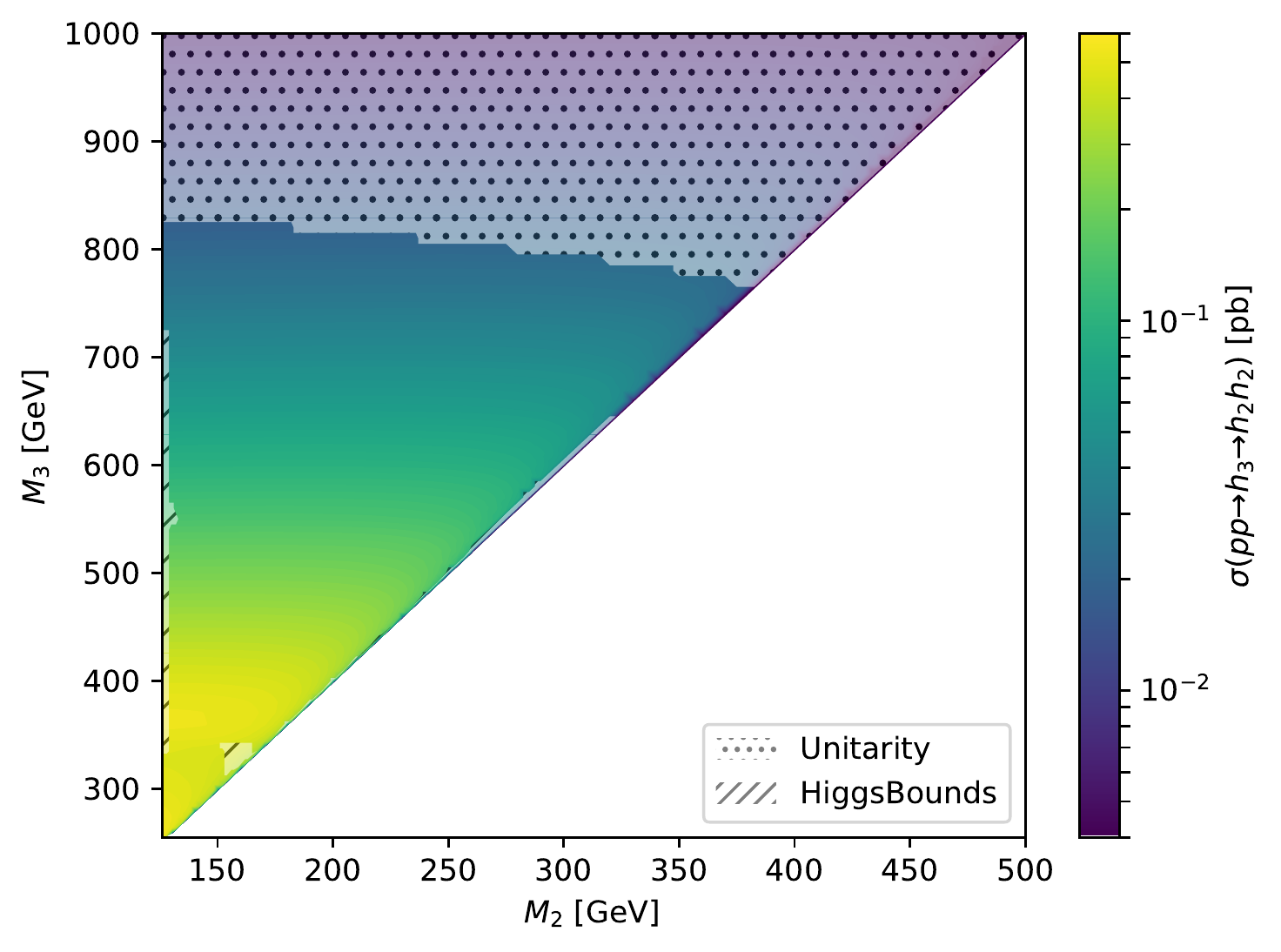}
    \caption{Benchmark plane \BP{6} for the decay signature $h_3 \to h_2h_2$
        with $h_{125}\equiv h_1$, defined in the $(M_2, M_3)$ plane. The color
        code shows $\BR{h_3 \to h_2h_2}$ (\emph{left panel}) and the signal rate
        for $pp \to h_3 \to h_2h_2$ (\emph{right panel}). The shaded region at
        large $M_3$ is excluded by perturbative unitarity. The shaded region at
        $M_2\sim\SI{125}{\GeV}$ is excluded by searches for resonant double
        Higgs production~\cite{Sirunyan:2018two}, and the shaded parameter
        region around $M_2\simeq \SI{160}{\GeV}$, $M_3\simeq \SI{330}{\GeV}$ by
        an ATLAS search for $h_3\to h_2 h_2 \to W^+W^-
        W^+W^-$~\cite{Aaboud:2018ksn} via \HiggsBounds.}\label{fig:bp6}
\end{figure}

\Cref{fig:bp6} shows the resulting $(M_2, M_3)$ parameter plane. The decay
channel $h_3\to h_2h_2$ --- shown in \cref{fig:bp6} (\emph{left}) --- is the
dominant decay mode of $h_3$ over the entire accessible parameter range with a
BR $\gtrsim \SI{75}{\%}$ except close to the kinematic threshold. The heavy
scalar $h_3$ is produced with about $\kappa_3^2=\SI{6}{\%}$ of the corresponding
SM predicted rate. The resulting signal cross section in \cref{fig:bp6}
(\emph{right}) reaches $\sim \SI{0.5}{\pb}$ in the low mass range, $M_3\lesssim
\SI{400}{\GeV}$, where $h_2$ decays directly to SM particles. The signal rates
in the mass range $M_3\gtrsim \SI{600}{\GeV}$, which is interesting for cascade
decays, can reach up to $\SI{100}{\fb}$ for $pp \to h_3 \to h_2h_2$ at the
$\SI{13}{\TeV}$ LHC\@. In \BP{6} the total width of of $h_3$ can reach up to
$\Gamma_3/M_3\sim\SI{14}{\%}$ without violating the unitarity constraint.
Therefore, it may be important to include finite width effects in experimental
analyses of this scenario.

The shaded region at large masses, $M_3 \gtrsim \SI{800}{\GeV}$, indicates that
the parameter region is in conflict with perturbative unitarity. Additionally,
experimental searches~\cite{Sirunyan:2018two} are beginning to probe the region
$M_2\sim\SI{125}{\GeV}$. Similar to the discussion of \BP{5}, this is a limit on
$h_3\to h_{125}h_{125}$ which is sensitive under the assumption that $h_2$ and
$h_1\equiv h_{125}$ cannot be experimentally distinguished if they are close in
mass. Moreover, a first ATLAS search for the signature $pp\to h_3 \to h_2 h_2
\to W^+W^-W^+W^-$~\cite{Aaboud:2018ksn} constrains a small region around
$M_2\simeq \SI{160}{\GeV}$, $M_3\simeq \SI{330}{\GeV}$, as shown in
\cref{fig:bp6}. We expect this search analysis to sensitively probe this
benchmark scenario in the future, in particular, if the currently considered
mass range is extended. The ATLAS search only considers $h_3$ masses up to the
$t\bar{t}$ threshold, $M_3 \le \SI{340}{\GeV}$. However, as we discuss here, the
$W^+W^-W^+W^-$ final state remains the dominant four-particle SM final
state even beyond the $t\bar{t}$ threshold.

\begin{figure}
    \centering
    \subfloat{\includegraphics[width=.49\linewidth]{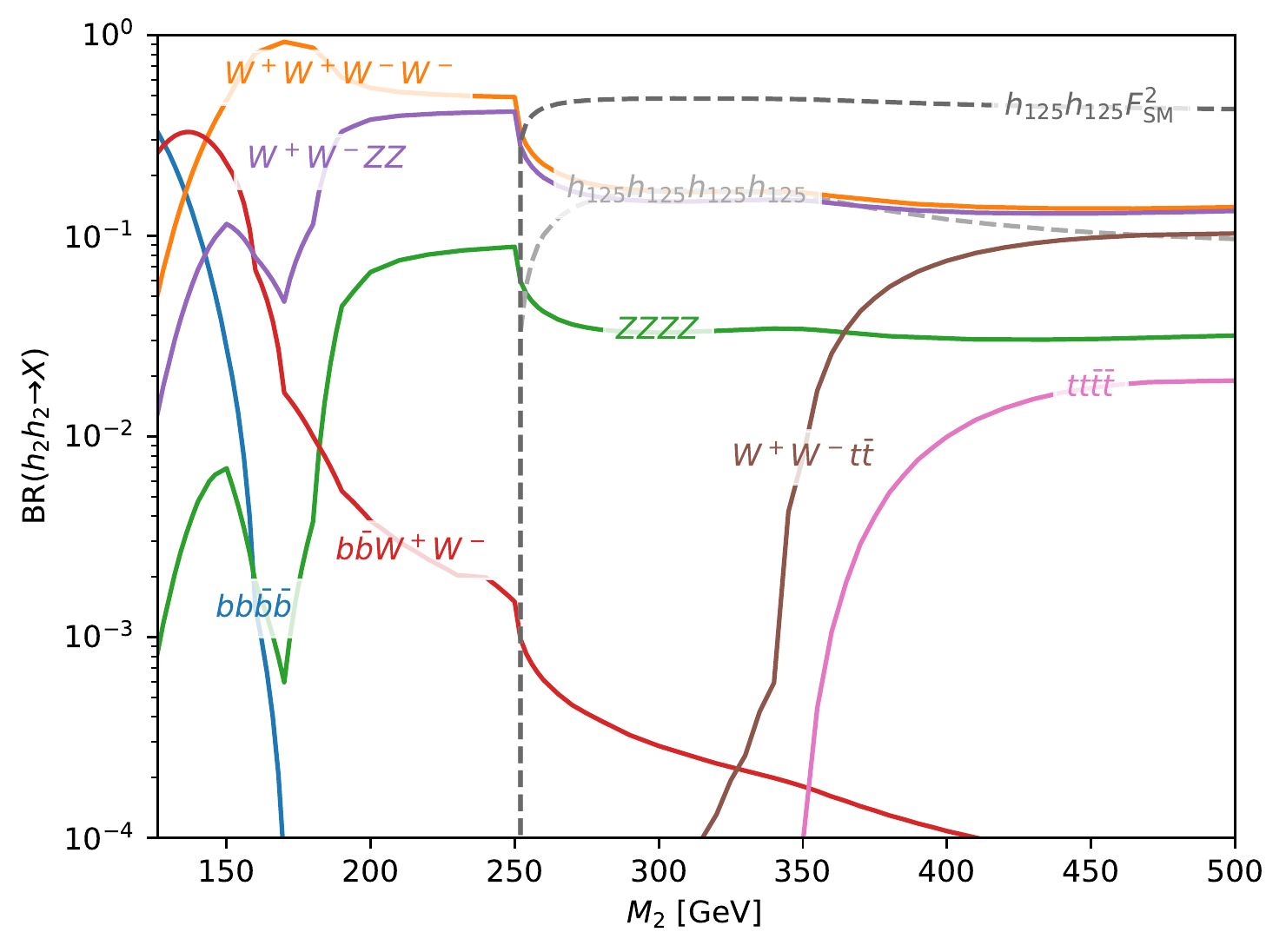}}
    \subfloat{\includegraphics[width=.49\linewidth]{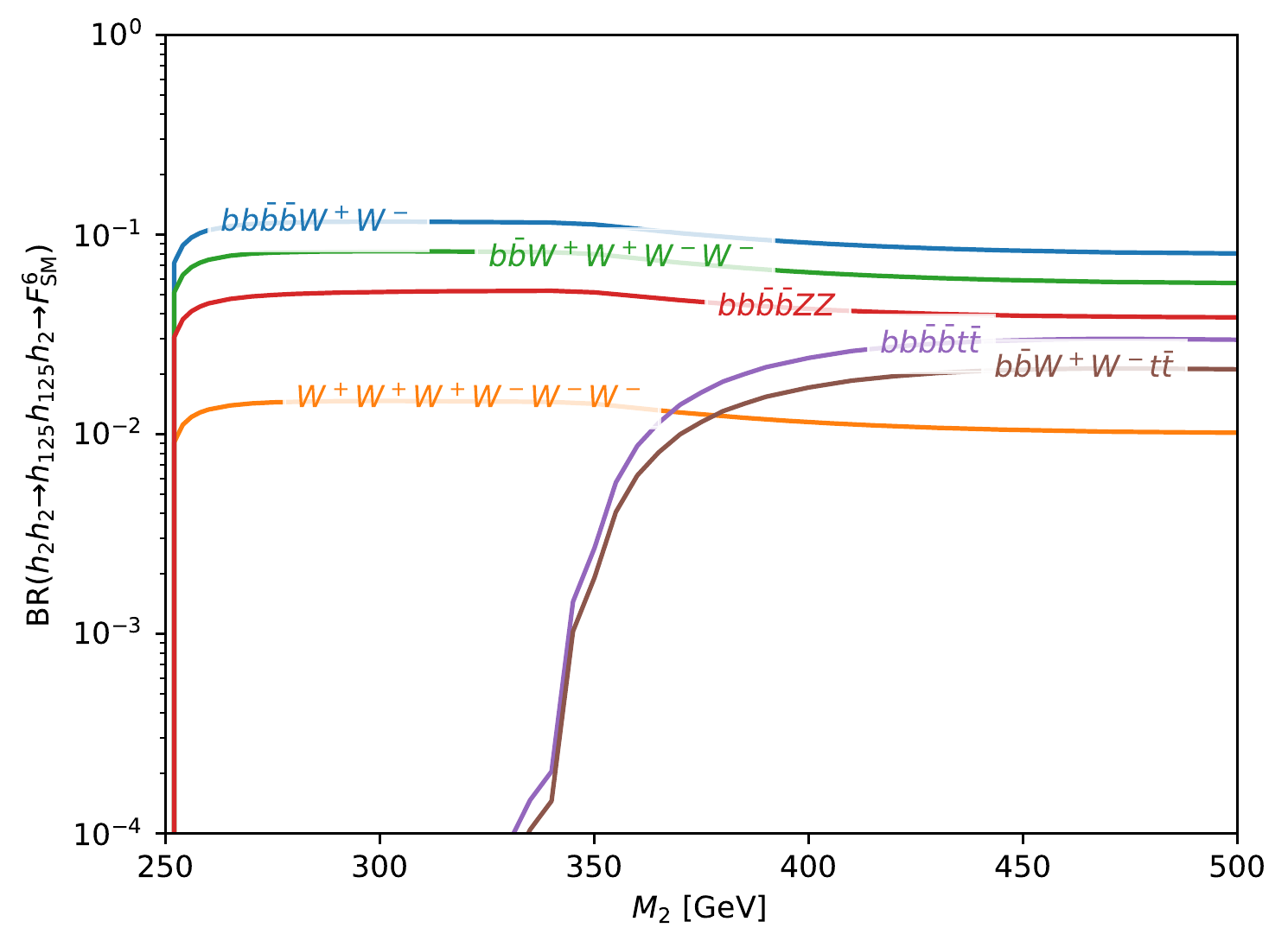}}
    \caption{Branching ratios of the $h_2h_2$ state of \BP{6}. The \emph{left panel}
    contains a selection of final states from direct decays of $h_2\to
        F_\text{SM}$ and (inclusive) decays involving $h_2 \to h_{125}h_{125}$ (both
    single and double cascade).  The \emph{right panel} shows the most
    important six particle SM final states, $F_\text{SM}^6$, that originate
    from a single cascade $h_2h_2\to h_{125}h_{125} F_\text{SM}$.
}\label{fig:bp6brs}
\end{figure}

\Cref{fig:bp6brs} shows the BRs of the decays of the $h_2h_2$ state resulting
from the $h_3$ decay in \BP{6}. At low $M_2 < \SI{250}{\GeV}$ only $h_2\to
F_\text{SM}$ decays are kinematically allowed. As shown in \cref{fig:bp6brs}
(\emph{left}), the dominant final state is $W^+W^-W^+W^-$, followed by
$b\bar{b}W^+W^-$ at low masses $M_2 \lesssim \SI{160}{\GeV}$ and $W^+W^-ZZ$ at
larger mass values.

For $M_2\gtrsim \SI{250}{\GeV}$ {the branching ratio for $h_2\to
h_{125}h_{125}$ is about $\SI{40}{\%}$ and} all three classes of decay chains
from \cref{fig:cascades} can occur in \BP{6}: direct decays of $h_2h_2\to
F_\text{SM}$; single cascade decays $h_2h_2\to h_{125}h_{125}h_2\to
F_\text{SM}$; and double cascade decays $h_2h_2\to
h_{125}h_{125}h_{125}h_{125}\to F_\text{SM}$, where the latter leads to a
spectacular final state composed of four $h_{125}$. The BRs for direct decays of
$h_2h_2$ to four-particle $F_\text{SM}$ are given in \cref{fig:bp6brs}
(\emph{left}). The dominant final states of this class are $W^+W^-W^+W^-$ and
$W^+W^-ZZ$, with $W^+W^-t\bar{t}$ becoming comparable at high $M_2$.
\Cref{fig:bp6brs} (\emph{left}) also shows the ``inclusive'' branching ratio for
the single cascade $h_{125}h_{125} F_\text{SM}$ (summed over all possible
$h_2\to F_\text{SM}$) and the double cascade decay rate to the
$h_{125}h_{125}h_{125}h_{125}$ final state.

The branching ratios of $h_2h_2$ into various six-particle SM final states via
the single cascade decay are shown in \cref{fig:bp6brs} (\emph{right}) as a
function of $M_2$. The most important decay modes involve $b$ quarks and $W$
bosons and --- due to combinatorial enhancement --- have decay rates comparable
to the four-particle final states. The decay $h_2h_2\to h_{125}h_{125}h_2\to
b\bar{b}b\bar{b}W^+W^-$ is the third most likely decay mode of $h_2h_2$ for
$\SI{250}{\GeV} < M_2 < \SI{350}{\GeV}$.

\begin{table}
    \centering
    \begin{tabularx}{\textwidth}{bssssssss}
        \toprule
        $\BR{X\to F^8_\text{SM}}$      & $6b\,2W$ & $8b$  & $4b\,4W$ & $4b\,2W\,2\tau$ & $6b\,2\tau$ & $4b\,2W\,2Z$ & $6b\,2Z$ & $6b\,2\gamma$ \\
        \midrule
        $h_{125}h_{125}h_{125}h_{125}$ & 17\%     & 12\%  & 9.2\%    & 5.5\%           & 5.2\%       & 2.3\%        & 2.2\%    & 0.19\%        \\
        $h_2 h_2$                      & 2.5\%    & 1.8\% & 1.3\%    & 0.8\%           & 0.75\%      & 0.34\%       & 0.31\%   & 0.027\%       \\
        \bottomrule
    \end{tabularx}
    \caption{{Decay rates of the $h_{125}h_{125}h_{125}h_{125}$ state in \BP{6},
    leading to a eight-particle SM final state, $F^8_\text{SM}$. The second row
    gives the corresponding rates originating from the $h_2h_2$ state, assuming}
    $\BR{h_2h_2\to
        h_{125}h_{125}h_{125}h_{125}}=\SI{14.5}{\%}$.}\label{tab:BP6quadbrs}
\end{table}

The branching ratios of the $h_{125}h_{125}h_{125}h_{125}\to F^8_\text{SM}$
decays via a double cascade are independent of the model parameters. They are
given in \cref{tab:BP6quadbrs}. Since the BR for the double cascade shown in
\cref{fig:bp6brs} (\emph{left}) is almost independent of $M_2$ we include in the
second row of \cref{tab:BP6quadbrs} an approximate branching ratio for the decay
of $h_2h_2$ into an eight-particle SM final state through the double cascade.
For this we use the averaged $\BR{h_2h_2\to h_{125}h_{125}h_{125}h_{125}} =
\SI{14.5}{\%}$, evaluated in the mass range $\SI{260}{\GeV} < M_2 <
\SI{500}{\GeV}$. The most important eight-particle final states are all
combinations of decays into $b$ quarks and $W$ bosons --- the most likely decay
products of $h_{125}$. Due to combinatorial factors their overall branching
fractions are, again, in some cases comparable to the four- and six-particle
final states. For example, the $b\bar{b}b\bar{b}b\bar{b}W^+W^-$ is similar to
the $ZZZZ$ final state rate for masses $M_3\sim\SI{300}-\SI{350}{\GeV}$. Near
the kinematic threshold, $M_3 \gtrsim \SI{500}{\GeV}$ and $M_2 \gtrsim
\SI{250}{\GeV}$, the signal cross section for $pp\rightarrow h_3\rightarrow
h_2h_2\rightarrow h_{125} h_{125} h_{125} h_{125}$ amounts to around
$\SI{14}{\fb}$.

\section{Conclusion}
\label{sec:conclusions}

We presented the collider phenomenology of a simple extension of the
SM Higgs sector, where two real scalar singlet fields are added to the particle
content. In this two-real-singlet model we imposed a discrete \Ztwo
symmetry for each scalar singlet field that is spontaneously broken by the
singlet field's vacuum expectation value. Consequently, all scalar fields
mix, leading to three neutral CP-even Higgs states $h_a$ ($a=1,2,3$). Any of
these states can be identified with the Higgs boson with mass $\simeq
    \SI{125}{\GeV}$ observed at the LHC\@.

The model leads to an interesting collider phenomenology for searches for the
additional Higgs states. Following the single production of one of the Higgs
states, $h_a$, this state can either decay directly to SM particles, or it can
decay into two lighter Higgs states, $h_a \to h_b h_c$, where the lighter states
can either be identical (``symmetric'' Higgs-to-Higgs decays with $b=c=1,2$), or
different  (``asymmetric'' Higgs-to-Higgs decays with $b=1$, $c=2$). In the
latter case, successive decays of the second lightest Higgs state to the
lightest Higgs state, $h_2 \to h_1 h_1$ may be possible if kinematically
allowed. This leads to interesting Higgs-to-Higgs cascade decay signatures, in
particular, $h_3 \to h_{1,2} h_2 \to F_\text{SM} h_1 h_1$ (``single cascade'')
and $h_3 \to h_2 h_2 \to h_1 h_1 h_1 h_1$ (``double cascade''), as shown in
\cref{fig:cascades}. We find that rates for all these possible Higgs-to-Higgs
decays can in general be sizable, easily dominating the direct decay modes to SM
particles.

Many of these Higgs-to-Higgs decay signatures have not been investigated
experimentally to date. We therefore presented six two-dimensional benchmark
scenarios to facilitate the design of dedicated experimental searches. Each
scenario is defined such that one of the novel signatures has a
nearly-maximal signal rate, while still obeying all theoretical and
experimental constraints on the model. Moreover, as the model can be
parametrized conveniently in terms of the relevant physical parameters, \ie
the three Higgs masses, three mixing angles (governing the Higgs coupling
strengths to SM particles) and the three vevs, the benchmark scenarios can
cover the entire kinematical phase space for the decay signatures, thus
rendering them as ideal references for experimental searches.

For each benchmark scenario, we discussed in detail the rates of the relevant
decays, as well as the expected signal rates in the TRSM at the
    \SI{13}{\TeV}~LHC\@. We furthermore provided an overview
of the most relevant SM particle final states, as a function of the relevant
mass parameters. This should provide a first step for experimental analyses to
estimate the discovery potential of corresponding searches. We expect that some
of the presented signatures can already be probed sensitively at the LHC  with
the data of $\sim \SI{150}{\fb^{-1}}$ per experiment collected during Run-II\@.

It should be kept in mind that the Higgs-to-Higgs decay signatures (and
potentially also the cascade decays)  discussed here can generically appear also
in other BSM models that feature three (or more) Higgs states. In that case,
however, the Higgs coupling properties do not necessarily agree with those of
the TRSM\@. This may result in different production rates of the
resonantly-produced Higgs state, as well as different decay rates, in particular
concerning the Higgs decays to SM particles. It is therefore important that
future experimental searches present their results as limits --- or ideally
measurements --- of the model-independent signal rate, as a function of the
relevant kinematical quantities (Higgs masses and, possibly, total widths).
Furthermore, Higgs-to-Higgs decays to possible SM particle final states that are
not dominant in the TRSM may still be worthwhile to probe experimentally, as the
anticipated rates may be different in other models. {In the case of a future discovery of an additional scalar state, signal rate measurements in various complementary production and decay modes will be crucial to probe its coupling structure and, in turn, to discriminate between the possible BSM interpretations.}

The exploration of the scalar sector --- leading to a better understanding of the mechanism of electroweak symmetry breaking --- is one of the most important scientific goals of the LHC program. This endeavor requires an open-minded and unbiased view on the potential collider signatures of new scalars. Our discussion of the TRSM and the presented benchmark scenarios demonstrate that there is a plethora of currently unexplored collider signatures involving Higgs-to-Higgs decays, and we hope that this work will initiate and facilitate the design of corresponding LHC searches in the near future.

\begin{acknowledgments}
    We thank the LHC Higgs Cross Section Working Group WG3 conveners for encouraging this study, as well as Massimiliano Grazzini and Georg Weiglein for useful comments. We also thank Claudia Seitz for helpful discussions regarding experimental questions and Liang Li for helpful explanations regarding the experimental search presented in Ref.~\cite{Aaboud:2018ksn}. The work of TS and JW is funded by the Deutsche Forschungsgemeinschaft (DFG, German Research Foundation) under Germany‘s Excellence Strategy -- EXC 2121 ``Quantum Universe'' -- 390833306. TR was supported by the European Union through the Programme Horizon 2020 via the COST action CA15108 - Connecting insights in fundamental physics (FUNDAMENTALCONNECTIONS), and additionally wants to thank the DESY Theory group for repeated hospitality while this work was completed.
\end{acknowledgments}
\appendix
\section{Comparison of the TRSM with the complex scalar singlet extension}
\label{app:translation}

The most general renormalizable and gauge invariant scalar potential of two real
singlet fields $S$ and $X$ and the SM Higgs doublet $\Phi$ is
\begin{equation*}
    \begin{aligned}
        V & = \mu_{\Phi}^2 \Phi^\dagger \Phi + \lambda_{\Phi} {(\Phi^\dagger\Phi)}^2
        + \mu_{S}^2 S^2 + \lambda_S S^4
        + \mu_{X}^2 X^2 + \lambda_X X^4                                                                              \\
          & \quad+ \lambda_{\Phi S} \Phi^\dagger \Phi S^2
        + \lambda_{\Phi X} \Phi^\dagger \Phi X^2
        + \lambda_{SX} S^2 X^2                                                                                       \\
          & \quad +a_S S+a_X X   + m_{SX} S X                                                                        \\
          & \quad+T_{SSS} S^3 + T_{XXX} X^3 +T_{SSX} S^2 X  + T_{SXX} S X^2                                          \\
          & \quad +T_{\Phi\Phi S}   \Phi^\dagger \Phi S +T_{\Phi\Phi X}  \Phi^\dagger \Phi X                         \\
          & \quad + \lambda_{SSSX} S^3 X +\lambda_{SXXX} S X^3  + \lambda_{\Phi \Phi SX} \Phi^\dagger \Phi S X\eqdot
    \end{aligned}\label{eq:potappa}
\end{equation*}
The first two lines correspond to the TRSM scalar potential, \cref{eq:TRSMpot},
while the remaining lines break the \Ztwo symmetries of \cref{eq:Z2syms}. These
21 real parameters relate to the 21 real parameters of the most general complex
singlet extension, Eq.~(1) of Ref.~\cite{Barger:2008jx} (using the same notation), via
\begin{align*}
    \mu^2_\Phi             & = \frac{m^2}{2}\eqcomma                                                   & \lambda_\Phi     & = \frac{\lambda}{4}\eqcomma                                                     \\
    \mu_S^2                & = \frac{1}{2}\left( b_2+b_1 \cos\phi_{b_1} \right)\eqcomma                & \lambda_S        & = \frac{1}{4}\left( d_1 \cos\phi_{d_1}+d_3 \cos\phi_{d_3}+d_2 \right)\eqcomma   \\
    \mu_X^2                & = \frac{1}{2}\left( b_2-b_1 \cos\phi_{b_1} \right)\eqcomma                & \lambda_X        & = \frac{1}{4}\left(  d_1 \cos\phi_{d_1}-d_3 \cos\phi_{d_3}+d_2  \right)\eqcomma \\
    \lambda_{\Phi S}       & = \frac{1}{2}\left( \delta_2+\delta_3 \cos\phi_{\delta_3}\right)\eqcomma  & \lambda_{\Phi X} & = \frac{1}{2}\left( \delta_2-\delta_3 \cos\phi_{\delta_3} \right)\eqcomma       \\
    \lambda_{SX}           & = -\frac{3}{2} d_1 \cos\phi_{d_1}+\frac{d_2}{2}\eqcomma                                                                                                                        \\
    a_S                    & = 2 a_1 \cos\phi_{a_1}\eqcomma                                            & a_X              & = -2 a_1 \sin\phi_{a_1}\eqcomma\vphantom{\frac{1}{2}}                      \\
    m_{SX}                 & = -b_1 \sin\phi_{b_1}\eqcomma\vphantom{\frac{1}{2}}                                                                                                                            \\
    T_{SSS}                & = \frac{1}{3}\left( c_1 \cos\phi_{c_1}+c_2 \cos\phi_{c_2} \right)\eqcomma & T_{XXX}          & = \frac{1}{3} \left( c_1 \sin\phi_{c_1}-c_2 \sin\phi_{c_2} \right)\eqcomma      \\
    T_{SSX}                & = -c_1 \sin\phi_{c_1}-\frac{c_2}{3}\sin\phi_{c_2}\eqcomma                 & T_{SXX}          & = -c_1\cos\phi_{c_1}+\frac{c_2}{3}\cos\phi_{c_2}\eqcomma                        \\
    T_{\Phi\Phi S}         & = \frac{\delta_1}{2}\cos\phi_{\delta_1}\eqcomma                           & T_{\Phi\Phi X}   & = -\frac{\delta_1}{2} \sin\phi_{\delta_1}\eqcomma                               \\
    \lambda_{SSSX}         & = -d_1 \sin\phi_{d_1}-\frac{d_3}{2} \sin\phi_{d_3}\eqcomma                & \lambda_{SXXX}   & = d_1 \sin\phi_{d_1}-\frac{d_3}{2}\sin\phi_{d_3}\eqcomma                        \\
    \lambda_{\Phi \Phi SX} & = - \delta_3 \sin\phi_{\delta_3}\eqdot\vphantom{\frac{1}{2}}
\end{align*}
Using these relations we could alternatively parametrize the TRSM  as a complex
singlet extension where
\begin{equation*}
    a_1 = c_1 = c_2 = \delta_1 = 0\eqcomma \quad
    \sin\phi_{b_1} = \sin\phi_{d_1} = \sin\phi_{d_3} = \sin\phi_{\delta_3} = 0\eqdot\label{eq:Z2conds}
\end{equation*}
The $U(1)$ symmetry imposed on the complex scalar in Ref.~\cite{Coimbra:2013qq}
requires $\delta_1 = \delta_3 = c_1= c_2 = d_1 = b_3 = a_1 = b_1 = 0$. This forms a special case of the TRSM where
\begin{equation*}
    \mu^2_X = \mu^2_S\eqcomma\quad
    \lambda_{\Phi S} = \lambda_{\Phi X}\eqcomma\quad
    \lambda_S = \lambda_X = \frac{1}{2} \lambda_{SX}\eqdot
\end{equation*}
In this model soft $U(1)$ breaking terms --- such as $a_1,b_1\neq0$ --- are
required to avoid a massless Goldstone boson. With these terms included, the
resulting model is no longer a special case of the TRSM\@.
\bibliography{main}

\end{document}